\documentclass[twocolumn]{aastex63}

\submitjournal{ApJ}
\shorttitle{Sgr B2 2018}
\shortauthors{Rogers et al.}
\usepackage{amsmath}

\begin{document}

\title{New Constraints on Cosmic Particle Populations at the Galactic Center using X-ray Observations of the Molecular Cloud Sagittarius B2}

\correspondingauthor{Field Rogers}
\email{frrogers@mit.edu}

\author[0000-0003-3791-0107]{Field Rogers}
\affiliation{Massachusetts Institute of Technology, 77 Massachusetts Avenue, Cambridge, MA 02139 USA}

\author[0000-0002-2967-790X]{Shuo Zhang}
\affiliation{Bard College, 30 Campus Road, Annandale-on-Hudson, NY 12504 USA }

\author[0000-0002-6404-4737]{Kerstin Perez}
\affiliation{Massachusetts Institute of Technology, 77 Massachusetts Avenue, Cambridge, MA 02139 USA}

\author[0000-0003-0724-2742]{Ma\"ica Clavel}
\affiliation{Univ.\ Grenoble Alpes, CNRS, IPAG, 38000 Grenoble, France}

\author[0000-0003-3864-6897]{Afura Taylor}
\affiliation{Massachusetts Institute of Technology, 77 Massachusetts Avenue, Cambridge, MA 02139 USA}

\begin{abstract}

The Sagittarius B2 (Sgr B2) molecular cloud complex is an X-ray reflection nebula whose total emissions have continued to decrease since 2001 as it reprocesses one or more past energetic outbursts from the supermassive black hole Sagittarius A*. 
The X-ray reflection model explains the observed time variability and provides a window into the luminous evolutionary history of our nearest supermassive black hole. 
In light of evidence of elevated cosmic particle populations in the Galactic Center, X-rays from Sgr B2 {are also of interest} as a probe of low-energy (sub-GeV) cosmic rays, {which may be responsible for an increasing relative fraction of the nonthermal X-ray emission as the contribution from X-ray reflection decreases}. 
Here, we present the most recent \textit{NuSTAR} and \textit{XMM-Newton} observations of Sgr B2, from 2018. 
These reveal small-scale variations within lower-density portions of the complex, including brightening features, yet still enable upper limits on X-rays from low-energy cosmic particle interactions in Sgr B2. 
We present Fe K$\alpha$ fluxes from cloud regions of different densities, facilitating comparison with models of ambient LECR interactions throughout the cloud.

\end{abstract}
\keywords{Sagittarius B2, Molecular Cloud, X-ray Reflection, Cosmic Ray, Galactic Center}


\section{Introduction} \label{sec:intro}

Centered at $\sim$$100$\,pc projected distance from the supermassive black hole Sagittarius A* (Sgr A*; \citet{Ghez08}) at the dynamic center of the Galaxy, and $\sim$$8$\,kpc from Earth, the molecular cloud (MC) Sagittarius B2 (Sgr B2) is the densest and most massive such object in the Central Molecular Zone (CMZ), a region that extends several 100\,pc from Sgr A* and contains $\sim$$10\%$ of the Galaxy's total molecular material \citep{Morris96}. 
X-ray observations of Sgr B2 have revealed strong Fe K$\alpha$ line emission at 6.4\,keV \citep{Koyama96,Murakami01,Koyama07,Inui09,Terrier10,Nobukawa11,Terrier18} as well as a hard continuum up to $\sim$$100$\,keV \citep{Sunyaev93,Revnivtsev04,Terrier10,Zhang15}. These features, which imply energetic, nonthermal interactions capable of ionizing the K-shell electrons of neutral Fe, have made Sgr B2 an object of interest for decades. The X-ray picture is further complicated by the time-varying nature of this emission. Since the peak flux was last observed in 2001, the Fe K$\alpha$ emission has decreased with every subsequent observation, down to $\sim$$20\%$ of the peak by 2013 \citep{Zhang15,Terrier18}. The hard continuum emission from the complex has correspondingly decreased, by $\sim$$50\%$ from 2003 to 2019 \citep{Kuznetsova21}.

In a simplified model {\citep{Lis90,Vicente97}}, Sgr B2 consists of a dense ($(3-9)\times 10^6$\,H$_2$\,cm$^{-3}$) star-forming core with radius $\sim$$2-4^{\prime\prime}$, or\ $\sim$$0.15-0.3$\,pc given the $\sim$$8$\,kpc distance to Sgr B2 \citep{Reid09}. The core is surrounded by an envelope of intermediate density ($10^4-10 ^5$\,H$_2$\,cm$^{-3}$) with radius 2.2$^{\prime}$, or $\sim$$5$\,pc, and an extended diffuse ($\sim$$10^3$\,H$_2$\,cm$^{-3}$) region with radius $\sim$$9.9^{\prime}$ or $\sim$$22.5$\,pc. The model reproduces the observed column density $N_H \sim 10^{24}$\,cm$^{-2}$ through the {central} core
and the total mass $ \sim$$6\times 10^6 M_{\sun}$ of Sgr B2. In reality, Sgr B2 has a more complicated structure including several subdominant cores \citep{Benson84,Sato00,Etxaluze13,Schmiedeke16} and an asymmetric overall gas distribution as revealed by images of cold dust \citep{Molinari11}.

In the X-ray reflection nebula (XRN) model, the Fe K$\alpha$ X-rays originate in the reprocessing of  external X-rays via K-shell photoionization and subsequent fluorescence of neutral Fe gas while the continuum emission arises from Rayleigh and Compton scattering \citep{Sunyaev93,Koyama96,Sunyaev98}. Reprocessing of X-rays originating in past flaring activity of Sgr A* is the  widely accepted explanation of the time-variable nonthermal emission, where the time-variability emerges as the flares pass in and out of the MC \citep{Sunyaev93}. Meanwhile, emission due to multiple scattering is expected from the densest regions even after the flare has exited the cloud \citep{Sunyaev98,Odaka11,Molaro16}. 

A short ($<$10-year) and bright event taking $\sim$$10-20$ years to traverse the cloud could explain the peak luminosity from the Sgr B2 core as well as the subsequent dimming (see \citet{Terrier18} and references therein). 
Though direct observation shows that Sgr A* is presently in a quiescent state \citep{Baganoff03,Wang13,Corrales20}, the XRN picture of Sgr B2 and other MCs in the CMZ reveals that Sgr A* has been brighter in the past few hundred years, with at least two short outbursts \citep{Ponti10,Clavel13,Ponti13,Churazov17a,Chuard18,Terrier18}. 

A portion of the X-ray emission from Sgr B2 could arise from low-energy ($<$1\,GeV, i.e.\ highly ionizing) cosmic-ray (LECR) electrons or protons, where the Fe K$\alpha$ line arises from K-shell ionization of neutral Fe and the continuum arises from Bremsstrahlung processes \citep{Valinia00,Yusef-Zadeh07,Dogiel09a,LECR}. 
The observed rates of hydrogen ionization in the Galactic Center (GC) region, which are in excess of local rates by a factor of $\sim$10, require models with elevated GC LECR populations relative to the local galactic environment \citep{Indriolo12,LePetit16,Oka19}. 
Neither LECR electrons nor protons can explain the full time-varying flux. The cooling time for $\sim$$100$\,MeV protons is longer than the observed timescale of the decrease \citep{Terrier10}, and the proton population corresponding to the hydrogen ionization cannot explain the Fe K$\alpha$ emission in the bright state \citep{Dogiel13}. Meanwhile, cooling of LECR electrons could explain the time variability \citep{Yusef-Zadeh13}, but the peak X-ray flux cannot be easily explained by LECR electrons alone \citep{Revnivtsev04}, requiring a highly-tuned model, e.g.\ higher metallicity in Sgr B2 than surrounding clouds \citep{Yusef-Zadeh13}. However, any steady-state LECR population in the cloud contributes a constant nonthermal X-ray flux, in addition to the XRN flux. 

Evidence of elevated cosmic-ray particle (CR) populations {across a broad energy range} in the vicinity of the CMZ \citep{Hess06,Yusef-Zadeh13,Zhang14,Hess16,Heywood19,ZhangS20,Ponti21,Huang21} motivates a discussion of particle accelerators in the GC and the role of the GC in the dynamics in the Galaxy at large. 
Despite decades of discussion \citep{Ptuskin81}, these are not presently emphasized in Galactic CR propagation models. 
The models (i.e.~GALPROP \citep{Strong07}, DRAGON \citep{Evoli17}) currently used to predict fluxes at Earth assume CR acceleration in supernova remnants (SNR) distributed throughout the Galactic disk and largely ignore the GC region. A portion of CRs at Earth could be due to extreme processes at the GC \citep{Cheng12,Jaupart18,Anjos20}, although other recent work suggests that CRs originating in the GC are not able to escape \citep{Huang21}. LECRs in the Galaxy are excellent tracers for sites of CR acceleration, due to their slow propagation and high rates of energy loss, and are excellent probes of cosmic-ray propagation models \citep{Liu21}.

Measuring X-ray flux levels enables setting upper limits on ionizing power from LECRs within a given region of Sgr B2 \citep{Valinia00,Dogiel09b,Dogiel09a,Terrier10,LECR,Dogiel13,Zhang15}. 
The flux levels alone can only produce upper limits due to uncertainties in both the time-varying contribution from primary XRN flares and the contribution from multiple scattering, which can vary with longer timescales (see, e.g., the discussion by \citet{Chernyshov18}). 
Limits on X-ray emission from LECR propagation in MCs can probe ambient LECR populations in the CMZ, providing valuable input for CR models, though the ability of LECRs to {traverse} dense MCs is highly model dependent. 
The interactions of LECRs within clouds is of particular interest due to the impact of the local ionization environment inside clouds on star formation \citep{Morlino15}.

In this paper, we present recent deep observations of Sgr B2 obtained in 2018 by {\em XMM-Newton} and {\em NuSTAR}. Section \ref{sec:obs} details the observations and data preparation. In Section~\ref{sec:spatial} we show the X-ray morphology of Sgr B2 while in Section~\ref{sec:spec} we present spectral analysis of the central region. In Section~\ref{sec:timing} we compare the 2018 flux with earlier data to discuss the continued decrease in X-ray reflection since 2001, {while in} Section~\ref{sec:lecr} we { present our main results on} upper limits on Fe K$\alpha$ emission from ambient LECR proton and electron populations in different regions of Sgr B2. Finally, in Section~\ref{sec:discussion} we discuss these results in the GC context.


\section{Observation and Data Reduction}\label{sec:obs}

\begin{deluxetable}{lrrrD}
\vspace{5mm}
\tablecaption{{\em NuSTAR} and {\em XMM-Newton} observations of Sgr B2.\label{tab:obs}}
\tablewidth{0pt}
\tablehead{
\colhead{Instrument} & \colhead{Observation} & \colhead{Start Time} & \colhead{Exposure } &\\
\colhead{} &\colhead{ID} & \colhead{(UTC)} & \colhead{(ks)}
}
\startdata
{\em XMM-Newton} 	& 0112971501 & 2001-04-01T00:25:11 & {17.8} \\ 
{\em XMM-Newton} 	& 0203930101 & 2004-09-04T02:53:45 & {40.0} \\ 
{\em XMM-Newton} 	& 0694640601 & 2012-09-06T10:56:15 & {35.6} \\ 
{\em XMM-Newton} & 0802410101 & 2018-04-02T00:59:38 & {86.3} \\ 
\hline 
{\em NuSTAR}  & 40401001002 &2018-04-10T12:01:09& {149.0} \\ 
\enddata
\tablecomments{The exposure is reported for the time intervals used in the analysis. For {\em XMM-Newton} observations, the {\em pn}-equivalent exposure is given.}

\end{deluxetable}

Table~\ref{tab:obs} lists the observations discussed in this work. We present new observations of Sgr B2, taken jointly by the {\em XMM-Newton} and {\em NuSTAR} X-ray observatories in 2018. For comparison with the 2018 data, we also use archival {\em XMM-Newton} observations of Sgr B2. 

\subsection{XMM-Newton Observations}\label{sec:xmm}

\begin{figure*}[tbh]
\fig{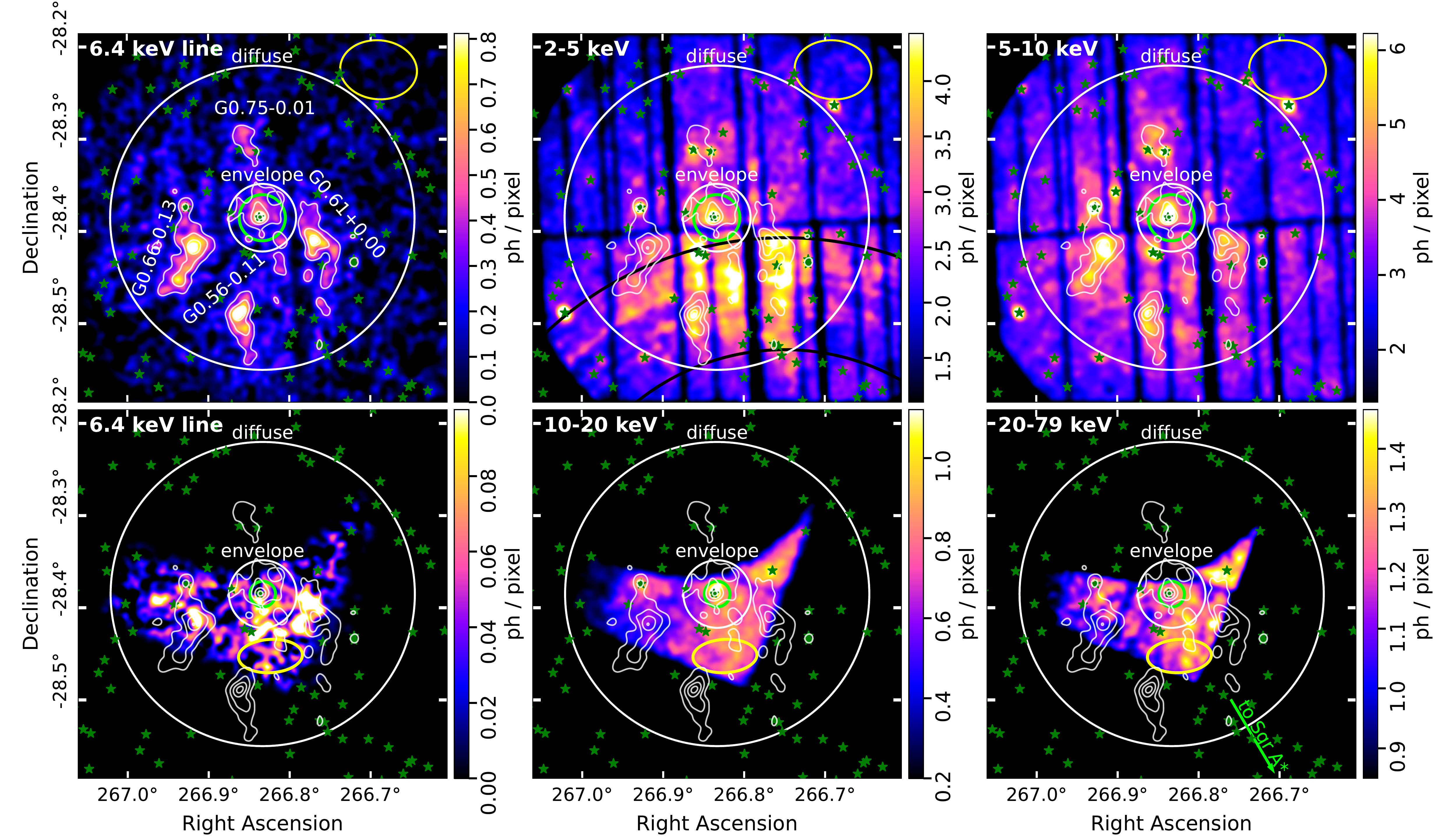}{\textwidth}{}
\caption{The 2018 X-ray morphology of the $24^{\prime} \times 24^{\prime}$ region surrounding Sgr B2 is shown as observed by {\em XMM-Newton pn} (top) in the 6.4\,keV line (left), $2-5$\,keV (center), and $5-10$\,keV (right) bands; and by {\em NuSTAR FPMA} (bottom) in the 6.4\,keV line (left), $10-20$\,keV (center), and $20-79$\,keV (right) bands. 
The 6.4\,keV line images are continuum subtracted as in Section~\ref{sec:spatial}. 
Contours (white) of the {\em XMM-Newton} 6.4\,keV map are overlaid on all images and illuminate the core and envelope of Sgr B2 as well as several substructures, labelled by their Galactic coordinates. 
The annular stray light pattern observed in all {\em EPIC} instruments is most evident in the $2-5$\,keV band (top center, black). The stray light in {\em FPMA} { from the bright off-axis source} is evident { in the curved} region removed from the top-left of the images, { while the diffuse stray light is apparent in the anomalously bright regions of the $10-20$\,keV and $20-79$\,keV bands}. Circles { (white)} indicating the diffuse (9.9$^{\prime}$ radius) and envelope (2.2$^{\prime}$ radius) regions of the simplified model are overlaid, while the core ($2-4^{\prime\prime}$ radius) is smaller than the angular resolution of both telescopes. The {locations of the} brightest ($>$$10^{-6}$\,ph\,cm$^{-2}$\,s$^{-1}$ in $2-7$\,keV) hard X-ray sources from the Chandra Source Catalog 2.0 \citep{Evans18} are shown (green stars), as well as the 90$^{\prime\prime}$ ({\em XMM-Newton}) and 50$^{\prime\prime}$ ({\em NuSTAR}) source regions {(lime)} and the respective elliptical regions used for background subtraction {(yellow)}. {The arrow (lime) indicates the direction to Sgr A*.} Color bars indicate intensity in photons per pixel. {The images have been smoothed using a 2D Gaussian kernel with standard deviation 3 pixels ({\em XMM-Newton pn}; pixel-size 4.3$^{\prime\prime}$) or 5 pixels ({\em NuSTAR FPMA}; pixel-size 2.5$^{\prime\prime}$).} 
\label{fig:spatial}}
\end{figure*}

{\em XMM-Newton} consists of three European Photon Imaging Camera (EPIC) instruments: two Metal Oxide Semiconductor (MOS) arrays and a pn array. These cameras detect X-rays from $0.15 - 15$\,keV with typical energy resolution of $\sim$$2-5\%$ and angular resolution of 6$^{\prime\prime}$ FWHM \citep{XMMmos,XMMpn}.

Analysis was performed using the {\em XMM-Newton} Extended Source Analysis Software (ESAS; \citet{Snowden08}) distributed with v.12.0.1 of the {\em XMM-Newton} Science Analysis Software. We reduced the data using the standard procedure and filtered the event files to exclude intervals affected by soft proton contamination.

Spectra were extracted with the ESAS {\em mos-spectra} and {\em pn-spectra} scripts. 
The MOS1 and MOS2 spectra were combined and all were rebinned with 3$\sigma$ significance after background subtraction. We analyzed MOS and pn data within the $2-10$ and $2-7.8$\,keV bands, respectively, where the pn spectra were truncated due to internal lines around 8\,keV.

\subsection{NuSTAR Observations}\label{sec:nustar}

{\em NuSTAR} operates in the $3-79$\,keV band using two focal plane modules ({\em FPMA} and {\em FPMB}) with angular resolution of 18$^{\prime\prime}$ (FWHM) and typical energy resolution of 400 eV (FWHM) at 10 keV \citep{NuSTAR13}. 

The data were reduced and analyzed using the NuSTAR Data Analysis Software (NuSTARDAS) v.1.3.1 and HEASOFT v.6.24 {\citep{Heasoft}}. They were filtered for periods of high instrumental background due to South Atlantic Anomaly passages and according to a database of bad detector pixels. 
{The data quality was impacted by stray light (unfocused photons arriving directly onto the detector at large off-axis angles) from both bright isolated sources and diffuse X-ray backgrounds. }
We removed the pixels contaminated by stray light 
 from bright isolated sources {using a geometrical model of the telescopes} following 
 \citet{Krivonos14}. 
{The {\em FMPB} observation was disregarded because the removed pixels covered the Sgr B2 region. }
{In {\em FPMA}, pixels were removed as close as $\sim$50$^{\prime\prime}$ from the center of Sgr B2. }
{In contrast to the bright isolated sources, stray light from diffuse backgrounds fills the entire detector area with a non-uniform pattern, the brightness of which limits the signal-to-noise. }
We {therefore} use {\em NuSTAR} spectra in the range of $10 - 20$\,keV, where the signal-to-noise ratio is highest.

\section{Morphology of X-ray Emission}\label{sec:spatial}

\begin{figure*}[tbh]
\fig{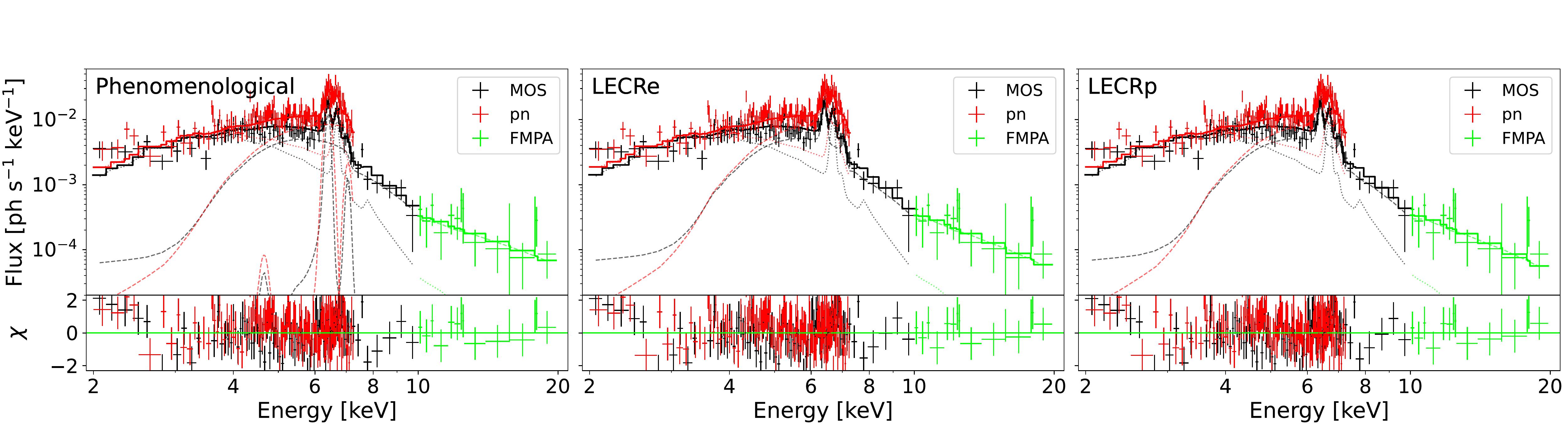}{\textwidth}{}
\caption{The {background-subtracted} 2018 spectra of the inner 90$^{\prime\prime}$ of Sgr B2 as observed by {\em XMM-Newton MOS} (black, $2-10$\,keV) and {\em pn} (red, $2-7.8$\,keV) and the inner 50$^{\prime\prime}$ as observed by {\em NuSTAR FPMA} (green, $10-20$\,keV) are fitted with the phenomenological model (left), the {\tt LECRe} model (center), and the {\tt LECRp} model (right). The {\em XMM-Newton} data are binned with $3\sigma$ {significance} while the {\em NuSTAR} data are binned with $1.5\sigma$ {significance}. The best fit is shown in the solid lines. {The contributions of the {\tt apec} (dotted) and the nonthermal spectral components (dashed; {\tt ga}, {\tt po}, {\tt LECRe}, and {\tt LECRp} for the respective models) are also shown}. 
All three models show satisfactory agreement with the data. {The excess below $\sim$2.3\,keV is compatible with foreground emission as also observed towards Sgr C (e.g.\ by \citet{Ryu13})}. 
\label{fig:spec}}
\end{figure*}

Figure~\ref{fig:spatial} presents the 2018 observations in the $24^{\prime} \times 24^{\prime}$ region centered on Sgr B2. The upper panel shows the {\em XMM-Newton pn} images in the continuum-subtracted 6.4\,keV line and in the $2-5$\,keV and $5-10$\,keV bands. 
The lower panel shows the {\em NuSTAR FPMA} images in the continuum-subtracted 6.4\,keV line and in the $10-20$\,keV and $20-79$\,keV bands. 
The 6.4\,keV line images were created by subtracting a continuum band, $5.8-6.2$\,keV, from a $6.2-6.6$\,keV signal band. 
The 90$^{\prime\prime}$ ({\em XMM-Newton}) and 50$^{\prime\prime}$ ({\em NuSTAR}) source regions used for the primary spectral analysis of the core and envelope are shown in dark blue, {where the smaller {\em NuSTAR} spectral extraction region is due to stray light contamination (see Section~\ref{sec:nustar})}. The background regions used for spectral analysis with each instrument are in green. The {\em XMM-Newton} background region is located outside the Sgr B2 complex. 
In contrast, the {\em NuSTAR} background region is located within the diffuse region of Sgr B2, due to the limited field of view {(see Section \ref{sec:nustar}). All regions shown in Figure~\ref{fig:spatial} are listed in Table~\ref{tab:regions} in the Appendix}.

The {\em XMM-Newton} 6.4\,keV line map shows that the core of Sgr B2 is detected at 13$\sigma$ significance within the envelope region. The core is also detected at $>$$5\sigma$ significance in {the $2-10$\,keV band of} {\em XMM-Newton} and by {\em NuSTAR} from $10-20$\,keV. In the $20-79$\,keV band, the {\em NuSTAR} observation is dominated by background, and the core is not significantly detected ($<$$3\sigma$).

In addition to the central core and envelope, Fe K$\alpha$ emission is detected at $>$$5\sigma$ significance from four substructures within the diffuse region of the Sgr B2 in the projected plane. Two of these substructures coincide spatially with the cloud features previously identified by \citet{Terrier18} as G0.66-0.13 and G0.56-0.11; \citet{Zhang15} also detected G0.66-0.13 in hard X-rays in 2013. 
Here we additionally report two substructures, labeled G0.61+0.00 and G0.75-0.01, which were not detected in previous observations. Of these four substructures, only G0.66-0.13 and G0.61+0.00 lie within the {\em NuSTAR} field of view. These substructures, which are fainter than the core, are detected by {\em NuSTAR} at 6.4\,keV but not resolved above background in the higher energy bands.


\section{Spectral Analysis of the Sgr B2 Core}\label{sec:spec}

\begin{deluxetable*}{LCChhhCCC}
\vspace{5mm}
\tablecaption{Best-fit spectral parameters are shown for a joint fit of the 2018 {\em XMM-Newton} and {\em NuSTAR} observations, using the central 90$^{\prime\prime}$ of Sgr B2 for {\em XMM-Newton} and central 50$^{\prime\prime}$ of Sgr B2 for {\em NuSTAR}. We report flux parameters for the 90$^{\prime\prime}$ region. \label{tab:params}}
\tablewidth{0pt}
\tablehead{
\colhead{Parameter} &\colhead{Unit}&\colhead{Phenomenological\tablenotemark{a}} &\nocolhead{MyTorus\tablenotemark{b}} &\nocolhead{Walls\tablenotemark{c}} &\nocolhead{CREFL16\tablenotemark{d}} &\colhead{LECRe\tablenotemark{b}} &\colhead{LECRp\tablenotemark{b}}  &\colhead{{LECRp (Z = 1)\tablenotemark{b}}} 
}
\startdata
N_H(f)& 10^{23}\,\textrm{cm}^{-2}&{0.9^{+0.2}_{-0.1}}&1.4\pm0.1&0.9^{+0.1}_{-0.2}&1.3\pm0.1&0.9\pm0.1&0.9\pm0.1& {1.0\pm0.1}\\
N_H(i)& 10^{23}\,\textrm{cm}^{-2}&{4.6^{+0.7}_{-0.6}}&9.0^{+6.0}_{-2.5}&7.9^{+3.7}_{-2.1}&12.3^{+7.9}_{-4.5}&5.2^{+1.2}_{-1.1}&5.0^{+0.4}_{-1.0}& {3.7\pm0.7}\\
Z/Z_\odot\textrm{ (apec)}& &2\textrm{*}&2\textrm{*}&2\textrm{*}&2\textrm{*}&2\textrm{*}&2\textrm{*}&{2\textrm{*}}\\
Z/Z_\odot \textrm{ (cloud)}& &...&...&...&1.0\textrm{*}&1.9^{+0.8}_{-0.4}&0.5^{+0.3}_{-...} & {1\textrm{*}}\\
kT& \textrm{keV}&{4.3^{+1.0}_{-0.7}}&6.5^{+0.8}_{-0.7}&4.3^{+1.1}_{-0.8}&5.4^{+0.6}_{-0.9}&4.3^{+1.1}_{-0.7}&4.3^{+1.1}_{-0.7}&{4.8^{+0.8}_{-0.9}}\\
F_{apec}~(2-10\textrm{\,keV})& 10^{-13} \textrm{\,erg\,cm}^{-2}\textrm{\,s}^{-1}&{5.6\pm0.3}&9.5\pm0.4&5.7\pm0.3&7.6^{+1.2}_{-0.9}&5.6\pm0.3&5.6\pm0.3&{6.4\pm0.03}\\
F_{6.4\,\textrm{keV}}& 10^{-6}\textrm{\,ph\,cm}^{-2}\textrm{\,s}^{-1}&{6.7\pm0.8}&...&...&...&...&...&{...}\\
\Gamma_{pl}& &2.0\textrm{*}&...&...&...&...&...&{...}\\
\Lambda& 10^{24}\textrm{\,H-atoms\,cm}^{-2}&...&...&...&...&5.0\textrm{*}&5.0\textrm{*}&{5.0\textrm{*}}\\
s& &...&...&...&...&3.2^{+0.8}_{-0.7}&2.9^{+1.6}_{-1.2}&{1.5^{*}}\\
E_{min}& \textrm{keV}&...&...&...&...&3.2^{+27.7}_{-2.2}&5600^{+54000}_{-5600} & {unconstrained}\\
N_{LECR}& 10^{-6}\textrm{\,erg\,cm}^{-2}\textrm{\,s}^{-1}&...&...&...&...&0.9^{+2.0}_{-0.8}&0.6^{+75.8}_{-0.5}&{0.17^{+0.04}_{-0.02}}\\
\textrm{{multiplicative} factor}& &{0.11^{+0.04}_{-0.03}}&0.19\pm0.06&0.11^{+0.02}_{-0.03}&0.15^{+0.02}_{-0.04}&0.12^{+0.06}_{-0.04}&0.12^{+0.06}_{-0.04} & {0.10\pm0.03}\\
\chi^2_{\nu}~\textrm{(d.o.f)}& &{1.07~(257)}&1.29~(257)&1.09~(256)&1.16~(257)&1.08~(255)&1.08~(255) & {1.11~(257)} \\
\enddata
\tablecomments{The goodness of fit is estimated by $\chi^2_{\nu}$ and the number of degrees of freedom is given in parentheses. The errors represent 90\% confidence. The fluxes and normalizations are for the 90$^{\prime\prime}$ region. The {multiplicative} factor relates the flux from the 50$^{\prime\prime}$ {\em NuSTAR} source region to the 90$^{\prime\prime}$ {\em XMM-Newton} region. 
}
\tablenotetext{a}{The phenomenological model is given by {\tt wabs*(apec+wabs*po+ga+ga)} in {\tt XSPEC}. It is characterized by foreground absorption {\tt wabs}, parametrized by the interstellar hydrogen column density $N_H(f)$, and by internal cloud absorption parametrized by column density $N_H(i)$. The thermal {\tt apec} component is characterized from metallicity $Z/Z_\odot$ which could not be constrained and was thus fixed at 2, the plasma temperature $kT$, and total flux contribution of $F_{\textrm{apec}}$. The lines, {\tt ga}, were fixed at 6.4\,keV (neutral Fe K$\alpha$) and 7.06\,keV (neutral Fe K$\beta$), with the Fe K$\beta$ flux fixed at 0.15 of Fe K$\alpha$. 
The continuum was modeled by a pure powerlaw {\tt po} characterized by the photon index $\Gamma_{pl}$. We fixed $\Gamma_{pl} = 2$ according to the best fit obtained from {\em NuSTAR} data alone, to avoid skewing by the higher-statistic {\em XMM-Newton} data. }
\tablenotetext{b}{The LECR models are given by {\tt wabs*(apec+wabs*LECR)} in {\tt XSPEC}, where {\tt LECR} is the electron ({\tt LECRe}) or proton ({\tt LECRp}) {\tt XSPEC} model \citep{LECR}. The {\tt LECR} models are characterized by the maximum path length $\Lambda$ of particles in the cloud, as well as the minimum energy $E_{min}$ of particles able to traverse the cloud, and the metallicity $Z/Z_{\odot}$ of the cloud. In addition to these cloud parameters, the power law index $s$ of the incident particle population is obtained in the fit. {The fit with the LECRp model is reported under two conditions: first with the  model parameters allowed to vary, and, second (labelled $Z=1$) with parameters constrained to physical values. } {As described in the text, these {\tt LECR} spectral fitting results do not clearly correspond to a physical ambient LECR scenario due to the assumptions inherent in the modeling and because the 90$^{\prime\prime}$ region is not physically motivated as the X-ray production region for ambient LECRs.}}
\tablenotetext{*}{Starred parameters were not allowed to vary in the fit.}
\end{deluxetable*}

Figure~\ref{fig:spec} shows the {background-subtracted} spectrum of the central region of Sgr B2 as observed in 2018, overlaid with the best fits {to three} spectral models. {All spectral fitting was performed using {\tt XSPEC} software \citep{Arnaud96}}. In this section, we detail {the spectral fitting to} models including a phenomenological model in Section~\ref{sec:pospec}, {three XRN} models in Section~\ref{sec:xrnspec}, and models of LECR-induced X-rays in Section~\ref{sec:crspec}. 

We extracted spectra from {\em XMM-Newton} in a 90$^{\prime\prime}$ source region, which includes the cloud's core and part of the envelope, consistent with \citet{Zhang15}. 
We used a local background region, which includes any diffuse 6.4\,keV emission from the larger GC region \citep{Uchiyama13}.
Constrained by stray light contamination {(Section~\ref{sec:nustar})}, we extracted {\em NuSTAR} spectra from the inner 50$^{\prime\prime}$ and from {\em FMPA} only. 
Due to the smaller {\em NuSTAR} field of view, background subtraction was performed using an ellipse located within the diffuse region of Sgr B2. The spectral extraction regions are illustrated in Figure~\ref{fig:spatial} and listed in Table~\ref{tab:regions}. 

Background subtraction is expected to account for the instrument background as well as diffuse emission from the GC region. Any faint point sources within the selected background region are also subtracted. Due to the different sky regions, flux from the diffuse region of Sgr B2 is subtracted from the {\em NuSTAR} but not {\em XMM-Newton} spectra, {leading to an underestimation of the emission from the core, which is corrected in the next paragraph}. No hard point sources were detected in either background region, so above 10\,keV, the contribution from both point sources and the diffuse GC emission is expected to be small. 

The {\em XMM-Newton} and {\em NuSTAR} spectra differ in source and background regions used for spectral extraction, in energy band, and in instrument characteristics. Here, the {\em XMM-Newton} data are used from 2 to 10\,keV to constrain most spectral characteristics. The {\em NuSTAR} spectrum is used from 10 to 20\,keV to constrain the spectral index of the continuum, which is observed to be consistent with previous measurements. To facilitate a simultaneous spectral fit, we introduce a multiplicative factor {relating} the overall normalization of the {\em NuSTAR} spectrum to that of the {\em XMM-Newton} spectrum {as a free parameter in the fit}. 
We estimate a multiplicative factor of $\sim$0.16. 
Of this, a factor of $0.49\pm0.05$ in the relative normalization is attributed to the {smaller {\em NuSTAR} source region, where the factor, which depends on the distribution of emission in the source area, was calculated based on image analysis of the 6.4\,keV band of {\em XMM-Newton}. Based on similar image analysis, the location of} the {\em NuSTAR} background region {within the diffuse region of Sgr B2} contributes a factor of $0.66\pm0.11$. {Finally,} decrease of up to $\sim$$50\%$ is anticipated due to the larger {\em NuSTAR} point spread function.

In all models discussed below, we use {\tt apec} to model thermal emissions remaining after background subtraction, following \citet{Zhang15}. Other  works used two {\tt apec} components \citep{Muno04,Walls16}, where a cooler component at $1-2$\,keV accounts for diffuse GC X-ray emission while a warmer component at $6-8$\,keV accounts for unresolved point sources. Here, we use a single {\tt apec}, for direct comparison with results in \citet{Zhang15}. There were no significant differences in nonthermal model parameters between our reported results using the single {\tt apec} model and fits using two {\tt apec} components with fixed temperatures.

In Sections~\ref{sec:xrnspec}$-$\ref{sec:crspec} we also consider fitted metallicity {$Z$ of the cloud relative to solar abundance $Z_{\odot}$} as a metric for the physicality of a fit. In the CMZ, we expect $Z/Z_{\odot}$ in the range of $1-2$, based on previous measurements \citep{Revnivtsev04,Nobukawa11,Jones11}. We {nominally} assume $Z/Z_{\odot} = 2$ {in the fitting} but consider $1 \leq Z/Z_{\odot} \leq 2$ as reasonable.

\subsection{Phenomenological Model}\label{sec:pospec}

Throughout this paper, we use a phenomenological model to directly evaluate the 6.4\,keV line flux. 
This model is detailed by \citet{Zhang15} and given by {\tt wabs*(apec+wabs*po+ga+ga)}. The powerlaw continuum ({\tt po}) and the neutral Fe K$\alpha$ (6.4\,keV) and K$\beta$ (7.06 keV) lines ({\tt ga}) 
expected in both the X-ray reflection and LECR scenarios are included explicitly. 
The model also accounts for thermal plasma ({\tt apec}) emission that persists after background subtraction, as well as both intrinsic and foreground absorption ({\tt wabs}). {We} use the {\tt wabs} model, rather than updated models such as {\tt tbabs}, to facilitate direct comparison with previous works.

We fix the line energies at 6.40\,keV and 7.06\,keV, {noting that best-fit centroid energy of the Fe K$\alpha$ line is at ($6.40\pm0.01$)\,keV when it is allowed to vary. We also fix} the line widths at 10\,eV, i.e.\ much less than the energy resolution of the instruments, and constrain the Fe K$\beta$ normalization at K$\beta$/K$\alpha = 0.15$ \citep{Murakami01}.  We fit the {\tt apec} plasma temperature but fix the {\tt apec} metallicity $Z/ Z_{\odot} = 2$ because it is not constrained by the data. 
{The choice of $Z/ Z_{\odot}$ impacts the relative flux of the fitted {\tt apec} and {\tt po} parameters. } 
The intrinsic and foreground hydrogen column densities, $N_H(i)$ and $N_H(f)$, in {\tt wabs}, are also fitted. After obtaining spectral index $\Gamma\sim2$ from the {\em NuSTAR} data only, consistent with previous measurements \citep{Terrier10,Zhang15}, we fix $\Gamma=2$ in the combined fit. Following \citet{Zhang15,Hailey16}, fixing $\Gamma$ according to {\em NuSTAR} data prevents the higher statistics of the {\em XMM-Newton} data from skewing $\Gamma$ based on an energy region where the power law is both degenerate with the thermal emission and more strongly absorbed.

Figure~\ref{fig:spec} (left) shows the spectral fit for the inner 90$^{\prime\prime}$, and the best fit model parameters are in Table~\ref{tab:params}. We obtained a satisfactory fit with $\chi^2_{\nu}$ = {1.07 for 257 d.o.f.} 
The best fit foreground column density, $N_H(f) = {0.9^{+0.2}_{-0.1}}\times10^{23}\,\textrm{cm}^{-2}$, was {higher than} the expected value of $0.7\times10^{23}\,\textrm{cm}^{-2}$ to the GC, {and} the fitted intrinsic column density $N_H(i) = {4.6^{+0.7}_{-0.6}}\times10^{23}\,\textrm{cm}^{-2} $ was {comparable} to the best fit of  $5.0\pm1.3\times10^{23}\,\textrm{cm}^{-2}$ {found with {\em NuSTAR} for the same source region} \citep{Zhang15}. 
We do not expect a physical value for $N_H(i)$ in this case because it only represents an average over the region, rather than the complex scattering dynamics in the cloud. 
We note that all nonthermal parameters are consistent between the values reported here and those obtained in a fit with a two-{\tt apec} plasma. 

The flux of Fe K$\alpha$ photons, reported as $F_{6.4\textrm{\,keV}} = {(6.7\pm0.8)}\times 10^{-6}$\,ph\,cm$^{-2}$\,s$^{-1}$, depends only weakly on the other model parameters. 
In particular, {even with $N_H(f)$ and $N_H(i)$ frozen outside of their 1$\sigma$ best-fit ranges or $Z_{\textrm{apec}}$ and $\Gamma$ frozen at values other than those in Table~\ref{tab:params}, spectral fits with $0.03 \times10^{23}\textrm{\,cm}^{-2}< N_{H}(f)<2.0\times10^{23}\textrm{\,cm}^{-2}$, $3.5 \times10^{23}\textrm{\,cm}^{-2}< N_{H}(i)<11.0\times10^{23}\textrm{\,cm}^{-2}$, $1<Z/Z_{\odot} \textrm{(apec)}< 5$, and $1< \Gamma< 4$ yield values of $F_{6.4\textrm{\,keV}}$ consistent within statistical uncertainty to that reported in Table~\ref{tab:params} despite the degradation in the fit statistic}.

\subsection{X-ray Reflection Nebula Models}\label{sec:xrnspec}

We use three self-consistent {\tt XSPEC} models of X-ray emission in the XRN scenario, including the {\tt MyTorus} model \citep{Murphy09,MyTorus}, the model developed by \citet{Walls16}, and the uniform Cloud REFLection of 2016 ({\tt CREFL16}) model \citep{Churazov17}, to assess the compatibility of the 2018 emissions with an XRN origin. 
All three models produce acceptable fits, but most model parameters are poorly constrained. Therefore, we conclude only that the emission spectrum observed in 2018 is not inconsistent with a primarily XRN origin. 
Details of the fitting with the three spectral models are in Appendix~\ref{app:xrnspec}.

\begin{figure*}[tbh]
\fig{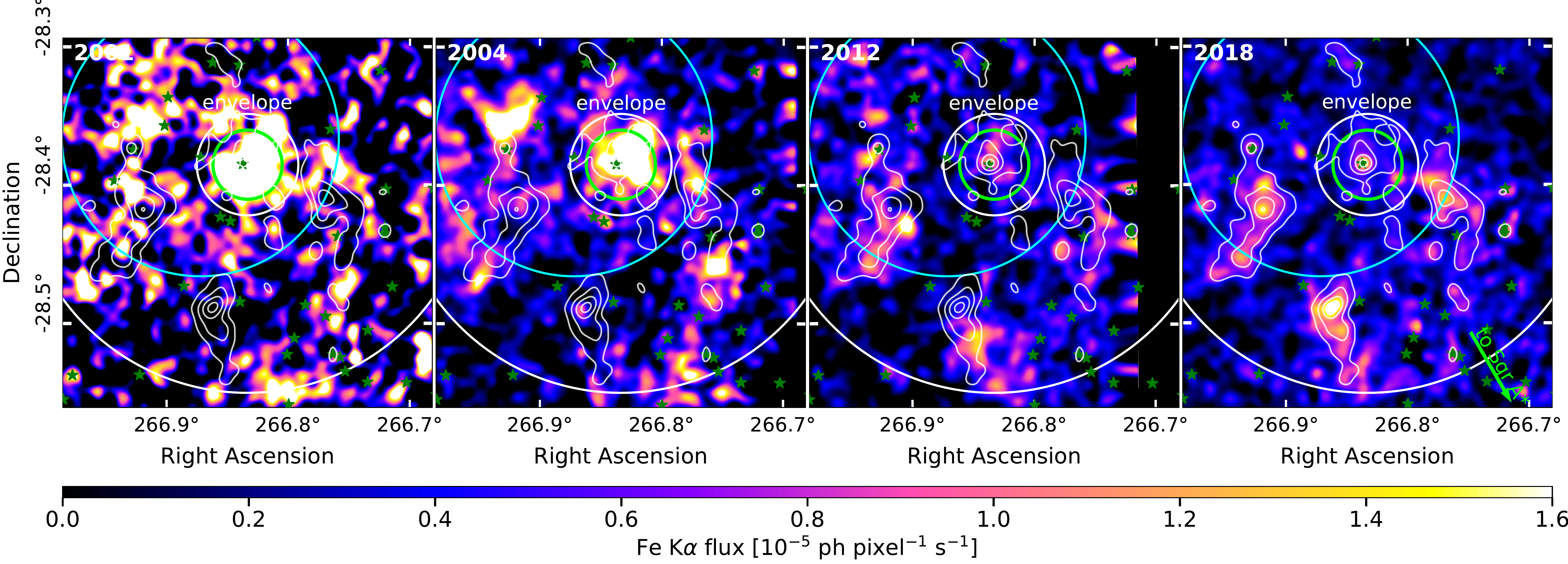}{\textwidth}{}
\caption{The morphology of Fe K$\alpha$ emission from Sgr B2 is shown as observed in 2001 (left), 2004 (center left), 2012 (center right), and 2018 (right) by {\em XMM-Newton pn}. The images are exposure corrected, with continuum subtraction performed as in Figure~\ref{fig:spatial}. The contours (white) from Figure~\ref{fig:spatial} illustrate the non-exposure-corrected Fe K$\alpha$ morphology observed in 2018. {The 6$^{\prime}$ (cyan) and 90$^{\prime\prime}$ {(lime)} regions corresponding to Section~\ref{sec:timingCentral} and Figure~\ref{fig:cloudDecrease} are shown.} {The 9.9$^{\prime}$-radius diffuse and the 2.2$^{\prime}$-radius envelope regions of {the simplified Sgr B2 gas model} (white circles)}, as well as the brightest hard X-ray sources (green stars) from the Chandra Source Catalog 2.0 \citep{Evans18}, are also shown for reference. The arrow (lime) points toward Sgr A*.
\label{fig:contours}}
\end{figure*}

\subsection{Low-energy Cosmic Ray Models}\label{sec:crspec}

We use the {\tt LECRe} and {\tt LECRp} {\tt XSPEC} models \citep{LECR} to understand if the spectral characteristics of the {Sgr B2 core in} 2018 are consistent with a LECR origin. {However, we note that the physics of LECR diffusion into dense clouds is highly model dependent (see Section~\ref{sec:ambient_cr}). {The {\tt LECR} ({\tt LECRe} and {\tt LECRp}) models were developed for the Arches cluster, which is smaller and less dense than Sgr B2 and features a stellar cluster that could accelerate CRs.} {Further, the 90$^{\prime\prime}$ region is not physically motivated as the X-ray production region corresponding to an ambient LECR population. Therefore, the results of the model fitting in this section do not correspond to the most physical ambient LECR scenario. More robust limits on ambient LECR populations are related to the Fe K$\alpha$ line fluxes from different cloud regions, reported in Section~\ref{sec:lecr}.}

The {\tt XSPEC} model is given by {\tt wabs*(apec+wabs*LECR)}, where the {\tt wabs} and {\tt apec} components account for foreground and internal absorption and any plasma emissions, as in Section~\ref{sec:pospec}.  
The {\tt LECR} models assume a MC is bombarded by CRs from an external source whose spectrum follows a powerlaw with index $s$. The remaining parameters, including the path length $\Lambda$ of CRs in the X-ray production (nonthermal) region, the minimum energy $E_{min}$ for a LECR to enter the X-ray production region, and the metallicity $Z$, are properties of the MC. 
The normalization $N_{LECR}$ describes the injected power {$\textrm{d}W/\textrm{d}t = 4\pi D^2 N_{LECR}$} by LECRs from $E_{min}$ to 1\,GeV, given distance $D$ to the MC.

The fitted parameters for the {\tt LECRe} and {\tt LECRp} models are in Table~\ref{tab:params} and fitted spectra are in Figure~\ref{fig:spec}. $\Lambda$ could not be constrained by the {data} and is frozen at $5\times10^{24}$ H-atoms per cm$^2$ in accordance with the column density through the core {following \citet{LECR,Zhang15}}. The metallicity in the {\tt LECR} model is that of the {molecular gas}, distinct from that of the {\tt apec} component.

In the electron ({\tt LECRe}) case, the fit is satisfactory with $\chi^2_{\nu} = 1.08$ for 255 d.o.f. 
The best-fit foreground $N_H(f) = (0.9\pm0.1) \times 10^{23}\,\textrm{cm}^{-2}$ and intrinsic $N_H(i) = 5.2^{+1.2}_{-1.1} \times 10^{23}\,\textrm{cm}^{-2}$ column densities are consistent with previous observations. The best-fit plasma temperature is $kT = 4.3^{+1.1}_{-0.7}$\,keV, and the cloud metallicity is $Z = 1.9^{+0.8}_{-0.4}$\,$Z_{\odot}$, consistent with the expected range of $1-2$\,$Z_{\odot}$. The fit favors no lower cutoff on LECR energies in the cloud, with $E_{min} = 3.2^{+27.7}_{-2.2}$\,keV, and an electron spectral index of $s = 3.2^{+0.8}_{-0.7}$. The fit normalization $N_{LECR} = 0.9^{+2.0}_{-0.8} \times 10^{-6}$\,erg\,cm$^{-2}$\,s$^{-1}$. {For $D =8$\,kpc to Sgr B2, this corresponds to} a limit (90\% confidence) on the power of LECR electrons, $\textrm{d}W/\textrm{d}t < 2.2 \times 10^{40} $\,erg\,s$^{-1}$ from the central 90$^{\prime\prime}$. {However, this upper limit is of limited utility given the assumptions in the {\tt LECR} model and considering that ambient LECRs are not expected to reach the core region of Sgr B2}. 

In the proton ({\tt LECRp}) case, the fit statistic for the central 90$^{\prime\prime}$ is also satisfactory, with $\chi^2_{\nu} = 1.08$ for 255 d.o.f. 
The best-fit foreground $N_H(f) = (0.9\pm0.1) \times 10^{23}\,\textrm{cm}^{-2}$ and intrinsic $N_H(i) = 5.0^{+0.4}_{-1.0} \times 10^{23}\,\textrm{cm}^{-2}$ column densities are consistent with previous observations. The plasma temperature is $kT = 4.3^{+1.1}_{-0.7}$\,keV.  $E_{min}$ is completely unconstrained, while the fit favors a similar LECR proton spectrum as in the {\tt LECRe} case, $s= 2.9^{+1.6}_{-1.2}$, consistent with the $1.5 < s< 2$ expected from diffusive shock acceleration \citep{LECR}. The cloud metallicity is fitted as $Z<0.8$\,$Z_{\odot}$ (90\% confidence), which is inconsistent with the expected value of $1-2$\,$Z_{\odot}$, so this fit does not represent a physical scenario. 
However by fixing the metallicity at $Z/Z_{\odot} = 1$ and {the proton power law index at $s = 1.5$, we} obtain a similar quality fit with $\chi^2_{\nu} = 1.11$ for 257 d.o.f.\ and physical model parameters  (see Table~\ref{tab:params}). In this case, $N_{LECR} = 1.7^{+0.4}_{-0.2}\times10^{-7}$\,erg\,cm$^{-2}$\,s$^{-1}$. {The corresponding} upper limit on the LECR proton power $\textrm{d}W/\textrm{d}t < 1.6\times 10^{39} $\,erg\,s$^{-1}$ {is subject to the same caveats as in the electron case above}.


\section{Time Variability of X-ray Reflection}\label{sec:timing}

Figure~\ref{fig:contours} shows the morphology of 6.4\,keV emission in four {\em XMM-Newton} observations of Sgr B2 from 2001 to 2018. The contours of the 2018 6.4\,keV emission are overlaid, for comparison with Figure~\ref{fig:spatial} despite the change in color scale. In these exposure-corrected and continuum-subtracted images, the surface brightness of the core and envelope decreases over time. In contrast to {this continued dimming}, substructures within the diffuse region, including those identified in Section~\ref{sec:spatial}, brighten and dim from one observation to the next.

Here, we discuss this changing Fe K$\alpha$ brightness and morphology from the cloud overall (Section~\ref{sec:timingCentral}) and from the substructures (Section~\ref{sec:timingDiffuse}). Details of the sky regions and spectral fitting are in Appendix~\ref{sec:timingdet} and Table~\ref{tab:regions}. All spectra were extracted using a local background region, such that the reported Fe K$\alpha$ flux is in excess of the diffuse emission reported by e.g.\ \citet{Uchiyama13}.

\begin{figure}[tbh]
\fig{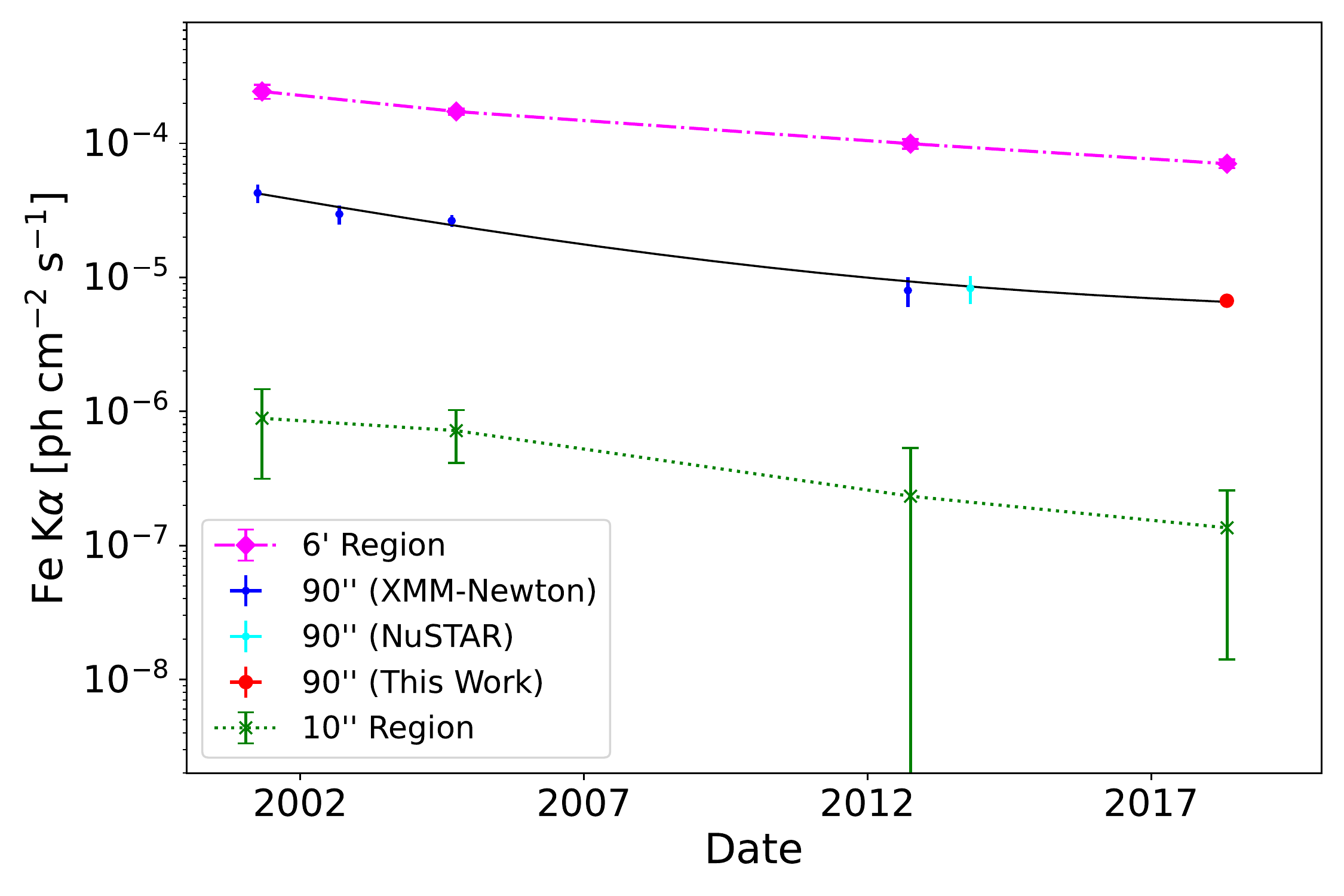}{0.45\textwidth}{}
\caption{
Time variability of the neutral Fe K$\alpha$ line flux is shown for a 6$^{\prime}$-radius region representing the cloud overall (magenta {dash-dotted}), the central 90$^{\prime\prime}$ (red, blue, and cyan) and the core {(green dotted)}. The light curve of the 6$^{\prime}$ region, which includes most of the diffuse emission, shows that the Fe K$\alpha$ flux from the cloud overall has decreased to $23\pm{3}\%$ of the 2001 {level} over this time period. 
The light curve for the central 90$^{\prime\prime}$ contains the data point calculated in Section~\ref{sec:spec} (red) alongside earlier measurements by \citet{Zhang15} with {\em XMM-Newton} (blue), and {\em NuSTAR} (cyan). 
The black curve is an exponential fitted to the data. 
The 2018 Fe K$\alpha$ flux from the inner 90$^{\prime\prime}$ of Sgr B2 is 
{$16\pm3\%$}
of the value measured in 2001. 
The green line shows the light curve for the central 10$^{\prime\prime}$-radius region, which corresponds to the $\sim$$15^{\prime\prime}$ half-power diameter of {\em XMM-Newton} together with the width of the core. 
\label{fig:cloudDecrease}}
\end{figure}

\begin{figure*}[tbh]
\fig{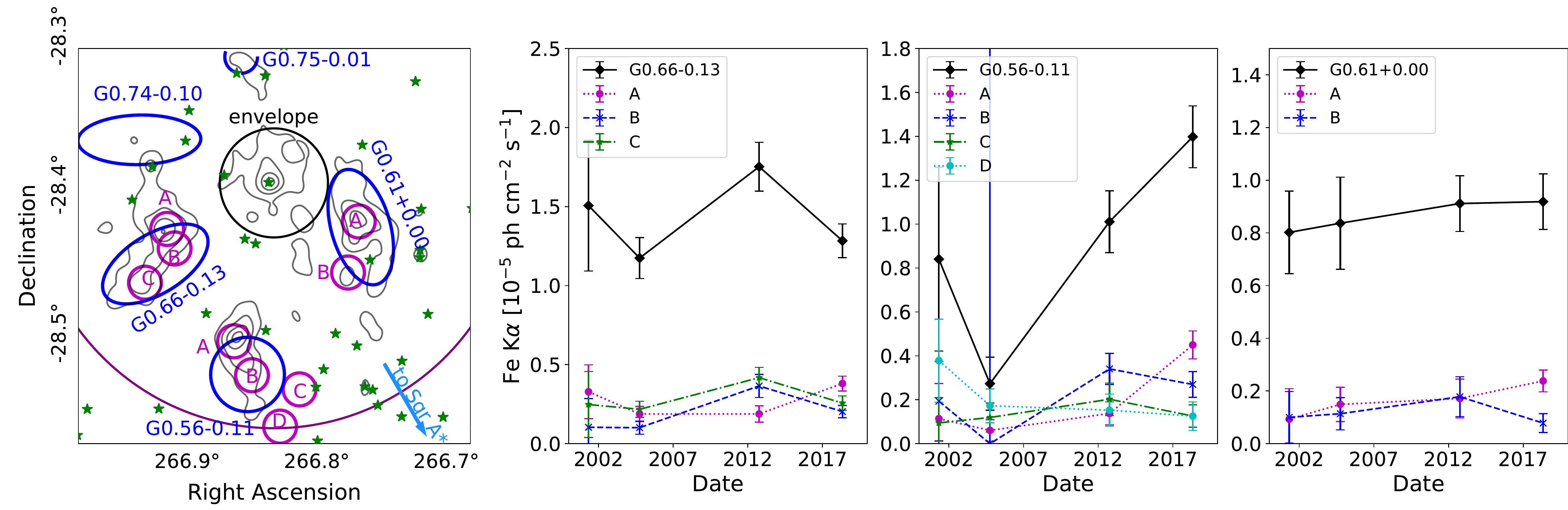}{\textwidth}{}
\caption{{\em Panel 1:} 
The contours of the 6.4\,keV line emission from 2018 are shown in the same projection as Figure~\ref{fig:contours}.
The regions (dark blue) identified as Sgr B2 substructures G0.66-0.13, G0.56-0.11, G0.75-0.10, and G0.61+0.00 are shown. 
G0.74-0.10, identified by \citet{Terrier18}, is also shown for reference, though it is not detected in 2018 and not treated in this work. 
We have further identified several 40$^{\prime\prime}$-radius features (magenta, A, B, C...) that illustrate the changes in morphology within each substructure over time. 
All region positions are given in Appendix~\ref{app:lecr_region_fits} Table~\ref{tab:regions}. 
The arrow (light blue) shows the direction to Sgr A*. 
{\em Panels 2-4:} 
The remaining panels show the Fe K$\alpha$ light curves extracted from the Sgr B2 substructures G0.66-0.13 (Panel 2), G0.56-0.11 (Panel 3), and G0.61+0.00 (Panel 4) outlined in Panel 1. The overall light curve from each substructure is in black, while the light curves of the corresponding 40$^{\prime\prime}$-radius features are plotted on the same axes. 
Substructure G0.66-0.13 was observed to brighten in 2012, but became dim again in 2018. Light curves for the 40$^{\prime\prime}$-radius regions (A, B, and C) associated with G0.66-0.13 behave differently over time, with circle A (magenta, farthest from Sgr A* in the projected plane) brightening only in 2018 while B and C follow the pattern of the parent substructure. 
Substructure G0.56-0.11 has brightened continuously since 2004, but analysis of 40$^{\prime\prime}$ features illustrates that this pattern is not uniform throughout the substructure. Instead, the brightening in 2012 is driven by region marked B (blue) while in 2018 the flux is driven by region A (magenta). 
Finally, while the light curve for the newly identified structure G0.61+0.00 is consistent with a constant in time, the 40$^{\prime\prime}$-radius features again illustrate an evolving morphology within the substructure.
\label{fig:substructures}}
\end{figure*}

\subsection{Time Variability of the Central Region}\label{sec:timingCentral}
Figure~\ref{fig:cloudDecrease} presents light curves of the Fe K$\alpha$ emission from Sgr B2, illustrating the behavior of the diffuse, envelope, and core regions.
We use a 6$^{\prime}$ region {(not concentric with the core; defined in Table~\ref{tab:regions} and illustrated in Figure~\ref{fig:contours})} to illustrate the behavior of the cloud over all. {We also analyze the 90$^{\prime\prime}$ region detailed in Section~\ref{sec:spec} to probe the behavior of the envelope, and a 10$^{\prime\prime}$ region to probe the core.}

{The 6$^{\prime}$ region} includes the Sgr B2 envelope and core and the bulk of the emission from the diffuse portion of the cloud. {Unlike the full 9.9$^{\prime}$ indicated by the simplified model, it is compatible with all four {\em XMM-Newton} observations.}
The resulting light curve illustrates that the total flux from Sgr B2 has continued to decrease with time, by a factor of $\sim$$3.5$ since 2001. {Relative to the 2012 level, the 2018 emission represents a $(29\pm8)\%$ decrease, indicating that the primary XRN component was contributing to the total nonthermal emission from this region at least as late as 2012. We note that \citet{Khabibullin21} reported the Fe K$\alpha$ emission from a comparable sky region using more recent (2019) commissioning data from {\it SRG/eRosita}. However, while the 2019 emission level is consistent with the 2018 level reported here, a higher-statistics {\it SGR/eROSITA} observation would be required to constrain the light curve. }
The 6$^{\prime}$ region directly corresponds to the sky region detected with the {\em INTEGRAL} observatory (which has a 6$^{\prime}$ resolution), facilitating comparison with the light curve of hard continuum emission reported by \citet{Kuznetsova21}. 

The light curve from the central 90$^{\prime\prime}$ consists of the Fe K$\alpha$ line flux measurement from Section~\ref{sec:pospec}, alongside measurements ($2001-2013$) from \citet{Zhang15}, demonstrating the compatibility of this work with previous measurements. 
Consistent with the images in Figure~\ref{fig:contours}, spectral analysis shows that the Fe K$\alpha$ line flux from the central 90$^{\prime\prime}$ (core and bulk of the envelope) decreased {over time}, with an overall decrease by a factor of {$6\pm1$ since 2001}. 
{The 2018 flux is $(80\pm20)\%$ of the 2012 flux, and the decrease since 2012 is not significant considering the statistical uncertainty, and we cannot determine based on the light curve if the flux from the central region has reached a constant level or if it will continue to decrease. } 

Finally, we show the light curve for the central 10$^{\prime\prime}$, which corresponds to the $\sim$$15^{\prime\prime}$ half-power diameter of {\em XMM-Newton} together with the width of the core. Though the fitting is less significant due to the small source size, we observe a similar pattern of decreasing emission  for the core as for the cloud overall. Based on the small relative flux contribution from the core, we conclude that behavior of the 90$^{\prime\prime}$ region is driven by interactions in the envelope.

{The true shape of the XRN light curve depends on the shape of the original flare and the details of the density profile of the cloud \citep{Sunyaev93}. Here, we fit the intensity $I$ of the emission from the central 90$^{\prime\prime}$ as a function of the time $t$ since January 1, 2001} 
 as an exponential decrease with a constant offset: . 
\begin{equation}
{I(t) = a*\exp(-t/\tau)+b}.
\label{eq:exp}
\end{equation}
{The best fit to Eq.~\eqref{eq:exp} yields normalization $a = {(3.9\pm0.5)} 
\times 10^{-5}$\,ph\,cm$^{-2}$\,s$^{-1}$, decay constant $\tau = {(5.2\pm 1.5)}$\,years, and constant flux $b = {(0.5\pm0.2)}
 \times 10^{-5}$\,ph\,cm$^{-2}$\,s$^{-1}$ {(1$\sigma$ confidence)}. 
{The best-fit $\tau$ is consistent with the expected superluminal light-crossing time for the  90$^{\prime\prime}$ region, indicating that the illuminating flare itself must have been short. The constant flux $b$ is inconsistent with 0, requiring an emission component that is stationary over the considered timescale. } 
 
In context of the {light curves}, Figure~\ref{fig:contours} illustrates how the geometry of the emission from the core and envelope has evolved over time. While the 2001 {map} is brightest on the Sgr A* side of the envelope, the emission is more balanced by 2004, and by 2012 and 2018, the envelope emission is more extended on the opposite side of Sgr A*, as illustrated by the contour lines. This provides further indication that the initial flare from Sgr A* has already passed through some or all of the envelope.

\subsection{Time Variability of Diffuse Substructures}\label{sec:timingDiffuse}

Figure~\ref{fig:substructures} (panel 1) shows the ellipsoid regions 
defined to correspond to the substructures identified in Section~\ref{sec:spatial}. 
The remaining panels in Figure~\ref{fig:substructures} show light curves of 6.4\,keV line emission from three of the substructures (G0.66-0.13, G0.56-0.11, and G0.61+0.00) that were significantly detected in 2018. 
The substructures behave differently over time, as G0.66-0.13 (black, second panel) brightens in 2012 and then dims again, while G0.56-0.11 (black, third panel) continues to brighten after 2004. 

Within each substructure, we defined several 40$^{\prime\prime}$-radius circular regions ($\sim$$10$-year light-crossing time) to illustrate the patterns of light that appear to be moving through the larger substructure.  
In each substructure, the circles are ordered from least negative declination (A, magenta, i.e.\ farthest from Sgr A* in the projected plane) to most negative, and the circle A brightens last. 
For G0.66-0.13, the light curves for circles B and C follow the same pattern as the parent structure, while circle A brightens in 2018, consistent with a flare originating at Sgr A* propagating through the cloud. 
We note that while these X-ray substructures were identified within the projected area of diffuse region of Sgr B2, we cannot exclude that they may correspond to other structures along the line of sight but outside of the Sgr B2 complex. Efforts to clarify the location of the substructures using line-of-sight velocity maps from the MOPRA 3\,mm survey \citep{Barnes15} were inconclusive.


\begin{figure}[tbh]
\fig{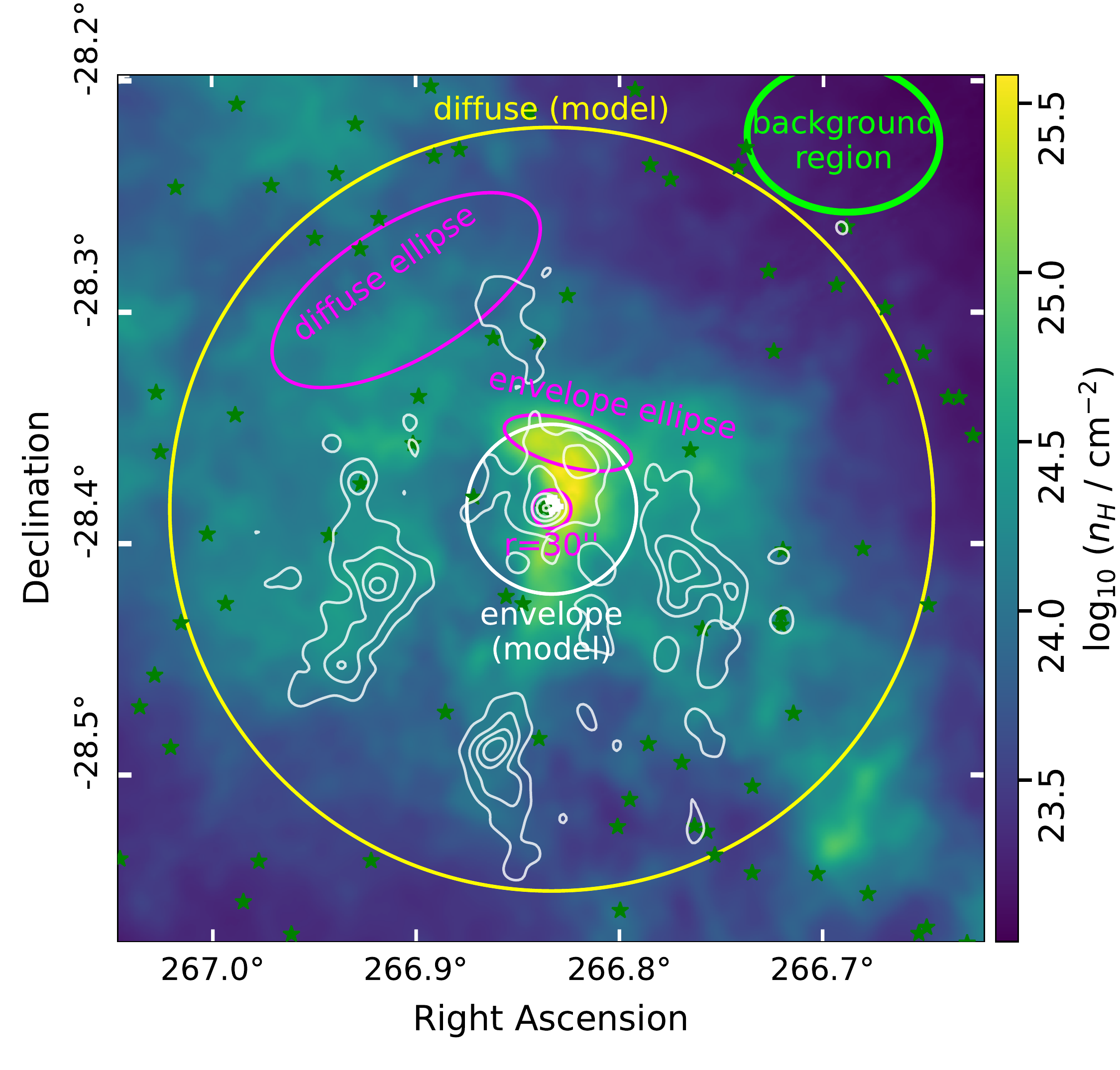}{0.45\textwidth}{}
\caption{
The hydrogen column density as measured by Herschel \citep{Molinari11} is shown in log scale and illustrates the complexity of the Sgr B2 structure compared to the simple model (yellow, 9.9$^{\prime}$ diffuse, and white, 2.2$^{\prime}$ envelope). The regions used in Table~\ref{tab:lecr_regions} are shown in magenta as the ellipses representing the clean diffuse and envelope regions, and with the {30$^{\prime\prime}$} circle. The background region is also shown (green ellipse), alongside the brightest hard point sources from the Chandra Source Catalog (green stars). 
\label{fig:nH}}
\end{figure}

\section{Low-energy Cosmic Ray Limits}\label{sec:lecr}

\begin{deluxetable*}{lLLLLD}
\vspace{5mm}
\tablecaption{The Fe K$\alpha$ flux and brightness from distinct regions of Sgr B2 can probe models of LECR transport into or production within Sgr B2.\label{tab:lecr_regions}}
\tablewidth{0pt}
\tablehead{
\colhead{Cloud Region}&\colhead{Fe K$\alpha$ flux}&\colhead{Fe K$\alpha$ surface brightness}&\colhead{$\chi^2_{\nu}\textrm{~(dof)}$}&\\
\colhead{}&\colhead{10$^{-6}$\textrm{\,ph\,cm}$^{-2}$\textrm{\,s}$^{-1}$}&\colhead{10$^{-7}$\textrm{\,ph\,cm}$^{-2}$\textrm{\,s}$^{-1}$\textrm{\,arcmin}$^{-2}$}&\colhead{}&
}
\startdata
Diffuse\textrm{*}&{6.1\pm 1.9}&{2.9\pm0.9}&{1.41~(23)}&\\
Env. (ellipse)$^{\dagger}$&{1.8\pm0.5}&{5.7\pm1.4}&{1.47~(41)}&\\
Env.\ ($0.5^{\prime} - 2.2^{\prime}$)$^{\dagger}$&{10.8\pm1.2}&{7.5\pm0.8}&{1.48~(317)}&\\
{Central 30$^{\prime\prime}$}&{1.3\pm0.3}&{16\pm4}&{1.25~(86)}&\\
\enddata

\tablecomments{Data are reported based on the 2018 {\em XMM-Newton} observation of Sgr B2. The region boundaries are in Appendix~\ref{app:lecr_region_fits} (Table~\ref{tab:regions}). Regions are circular or annular, with given angular size in radius, unless otherwise specified. Errors and upper limits indicate 90\% confidence. The corresponding spectral fits are in Figure~\ref{fig:lecr_fits}.}
\tablenotetext{*}{The region selected from the diffuse portion of the cloud is an ellipse, chosen to fall within the diffuse region in both the simplified model and the observed n$_H$ and to avoid hard point sources. The reported flux is thus the flux from this region, rather than the total flux from the diffuse region. Due to the more limited field of view of MOS1, we calculated flux using MOS2 and pn only for this region.}
\tablenotetext{\dagger}{Because the actual gas distribution in Sgr B2 is more complicated than the simplified model, two distinct sky regions were used to evaluate the envelope flux from Sgr B2, as shown in Figure~\ref{fig:nH}. The annular region, Env.\ (0.5$^{\prime} - 2.2^{\prime}$), represents the bulk of the envelope in the simplified Sgr B2 model. While the total flux measurement from this region may be useful, Sgr B2 has subdominant cores located within this annulus, and portions of this annular region have a column density more similar to the diffuse region, so the surface brightness should be interpreted with caution. On the other hand, the elliptical envelope region, Env.\ (ellipse), is a region with typical column density for the Sgr B2 envelope. While the flux for this region does not represent the total flux from the Sgr B2 envelope, the surface brightness is typical of the portions of the cloud with the intermediate column density associated with the envelope.}
\end{deluxetable*}

Theoretical efforts to model the propagation of LECRs into Sgr B2 rely on simplified models of the gas distribution and cloud dynamics. 
Models are necessitated by the complex gas structure, illustrated in Figure~\ref{fig:nH}. In this section, we have selected several sky regions {(illustrated in Figure~\ref{fig:nH} and detailed in Table~\ref{tab:regions}) that} are compatible with the diffuse, envelope, and core components of Sgr B2 in both the simplified model and the observed hydrogen column density while also avoiding the bright substructures identified in Section~\ref{sec:timingDiffuse}.

Table~\ref{tab:lecr_regions} presents the Fe K$\alpha$ flux and surface brightness based on spectral fitting of these sky regions. 
Details of the data and fitting are in Appendix~\ref{sec:lecrdet} {and Figure~\ref{fig:lecr_fits}}. 
The Fe K$\alpha$ surface brightness from the representative diffuse, envelope, and core regions of the cloud are {$(2.9\pm0.9)$}, {$(5.7\pm1.4)$}, and {$(16\pm4) \times 10^{-7}\,\textrm{ph}\,\textrm{cm}^{-2}\,\textrm{s}^{-1}\,\textrm{arcmin}^{-1}$}, respectively. 
{Since the 2018 data are the dimmest {deep} observations of Sgr B2 in this band to-date, these are the current best upper limits on Fe fluorescence due to ionization by LECRs.


\section{Discussion}\label{sec:discussion}

\subsection{Implications for the X-ray Reprocessing Scenario}\label{sec:discussion_XRN}

The core and envelope region, previously the brightest part of the cloud, with multiple cores detected \citep{Zhang15}, is now very faint in Fe K$\alpha$ fluorescence, as illustrated in Figures~\ref{fig:contours} and \ref{fig:cloudDecrease}. The brightest emission is restricted to the densest core. {The {pattern of} flux decrease from 2013 to 2018 from the core, envelope, and cloud overall indicates that most or all of the major X-ray flare} previously driving the overall luminosity has passed through the cloud by 2018. {The 2018 emission from the central 90$^{\prime\prime}$ is probably driven by a component that is not decreasing over at least the past several years. Such stationary or quasi-stationary behavior is expected from both multiple scattering and LECRs and could also arise from primary X-ray reflection if the short and bright flare responsible for the time variability was immediately followed by a longer-duration plateau with $\lesssim$0.1 of the peak luminosity. Reflection of the short and bright flare component is not ruled out as a subdominant contributor if it has not fully exited the 90$^{\prime\prime}$ region. Reflection of secondary flares in diffuse structures along the line of sight could also contribute.}

The total decrease in the Fe K$\alpha$ emission from the 6$^{\prime}$ region from 2004 to 2018 is consistent with the total decrease in the $30-80$\,keV continuum observed by {\em INTEGRAL} over the same sky region. The best-fit {\em INTEGRAL} light curve is a linear decay before 2011 and a constant level thereafter \citep{Kuznetsova21}. In comparison, Fe K$\alpha$ light curve reported here was decreasing at least as late as 2012, but due to limitations in the number and significance of the data points, the difference {between the two light curves} is not significant. 
A Fe K$\alpha$ light curve that is consistent with that of the hard continuum indicates that the emission mechanisms for the two energy scales are related. 
{This is expected, since the higher absorption at 6.4\,keV is partially compensated by multiple scatterings of the Fe K$\alpha$ line itself and because fluorescence events are induced by continuum photons above 7.1\,keV. If the emission is dominated by multiple scattering, the actual degree of correlation depends on geometry, optical depth and metallicity of the cloud \citep{Sunyaev98, Odaka11}.}

\citet{Kuznetsova21} suggest} multiple scattering of X-rays from the primary external flare as a {probable} origin of the $30-80$\,keV emission after 2011. 
{Multiple spectral handles can clarify the role of multiple scattering. First,} 
this signal is expected to be more pronounced in the morphology of the hard X-ray continuum than the fluorescent lines, as the absorption cross section is larger than the scattering cross section $\lesssim$10\,keV \citep{Odaka11,Churazov17}. Against a backdrop of fading emission, the 2013 {\em NuSTAR} detection \citep{Zhang15} of multiple cores above 10\,keV suggests that multiple scattering {already played} an increasingly important role. {Continued decrease of the Fe K$\alpha$ emission {once the} $30-80$\,keV emission {has reached a stationary level, as hinted by the data}, would support the multiple scattering origin of the $30-80$\,keV emission. 
{A future {\em NuSTAR} observation less contaminated by stray light would constrain the hard-continuum light curves from the core and envelope, allowing the cleanest probe of the multiple scattering scenario. While the 2018 observation was optimized to minimize stray light contamination, cleaner {\em NuSTAR} data could be obtained using a contiguous deep observation of a nearby off-source region to measure the stray light background, as in the 2013 observation \citep{Zhang15}.}

{In the Fe K$\alpha$ line, a Compton shoulder feature due to multiple scattering of fluorescence photons is expected on the low-energy side of the line complex \citep{Odaka11}.  
While the Compton shoulder is not resolvable with the energy resolution of {\em XMM-Newton}, the centroid of the 6.4\,keV line is expected to shift toward lower energies as the Compton shoulder becomes a more important contributor to the overall line flux \citep{Khabibullin21}. The fitted line centroid, $6.40\pm0.01$\,keV in 2018 for the central 90$^{\prime\prime}$,  is consistent with a constant value over the four {\it XMM-Newton} observations for both the 6$^{\prime}$ and 90$^{\prime\prime}$ regions. 
Future high-resolution X-ray spectrometers including {\em XRISM/Resolve} \citep{Ishisaki18} and {\em Athena/X-IFU} \citep{Barret18} could resolve the Compton shoulder, precisely measuring the relative contribution of multiple scattering \citep{Odaka11}.}

The morphological and brightness variation from several X-ray substructures {(reported in Section~\ref{sec:timingDiffuse}) external to the Sgr B2 envelope  also reveal implications for the timescale and number of the external X-ray flaring events}. 
The small bright regions, with their relatively short light crossing times, have a faster timescale of emission decrease from X-ray reprocessing and thus better reflect the timescale of the external source than the Sgr B2 envelope \citep{Terrier18}. In particular, G0.56-0.11, which was reported as brighting up in 2012 \citep{Terrier18}, is even brighter in 2018, with a significant morphologic change. While the 2012 emission is centered on clump B (see Figure~\ref{fig:substructures}), the 2018 emission is centered $\sim 13$ light years (projected distance) away, in clump A. The light curve of clump B gives an upper limit of $\sim$$14$ years for the timescale of the flare illuminating this region. 
Future observation by {\em XMM-Newton} could further constrain the timescales of these flares based on the future behavior of the ``A'' clumps from each substructure.

Similar light curves to G0.56-0.11 clump B are observed in G0.66-0.13 clumps B and C, suggesting that these two substructures may be illuminated by the same flaring event, if they are a similar distance from Sgr A*. 
Unfortunately, without knowledge of the line-of-sight positions and detailed $n_H$ distributions of these structures, we cannot clearly make this claim. 
\citet{Clavel13}, \citet{Chuard18}, and \citet{Terrier18} provided evidence for a minimum of two illuminating events propagating through the CMZ. 
In the case that these substructures are illuminated by the same event as the core, if we assume Sgr B2 to be at least 50\,pc in front of Sgr A* (following \citet{Walls16,Yan17}, and references therein), 
the substructures could be $\gtrsim$$60$\,pc behind Sgr B2, farther than the spatial extent of cloud. Therefore, if these substructures are linked to Sgr B2, they are illuminated by a secondary event.

\subsection{{Implications for} Low-energy Cosmic Rays}\label{sec:ambient_cr}

{This section discusses the upper limits on the Fe K$\alpha$ emission from different cloud regions presented in Section~\ref{sec:lecr} and Table~\ref{tab:lecr_regions} in the context of physical models of LECR interactions in Sgr B2. }

CR transport is modeled as a diffusive process modified by the effects of elastic and inelastic collisions, energy loss via ionization and excitation, and energy loss via bremsstrahlung and synchrotron radiation in the surrounding medium. {While transport into and within MCs is model dependent \citep{Gabici07,LECR,Gabici13,Dogiel15,Morlino15,Owen21}}, the low relative rates of hydrogen ionization observed within dense MCs {indicate} that LECRs do not freely traverse {these structures \citep{Oka05,VanderTak06,Indriolo12,Dogiel15}.}

Using a simplified {model} of Sgr B2, \citet{Dogiel15} calculate that LECR propagation in the envelope\footnote{\citet{Dogiel15} use the term `envelope' for our `diffuse region', and their `core' approximately corresponds to our `envelope'. They do not treat the dense star forming cores. For clarity, we have translated their terminology to match this work.} is best described by diffusion on turbulent magnetic fluctuations \citep{Dogiel87,Istomin13}. Meanwhile, diffusion is negligible in the diffuse region, where fluctuations are small. 
Considering ionization and excitation losses, protons (electrons) with kinetic energy $E \gtrsim 20$\,MeV (1\,MeV) traverse the diffuse region to reach the envelope, where they {are }absorbed within $0.1-0.3$\,pc \citep{Dogiel09b}. 
{The ambient LECR proton spectra derived by \citet{Dogiel15} would deposit the bulk of their energy in the Sgr B2 envelope.}

In the LECR proton case, \citet{Dogiel15} use the hydrogen ionization rates in Sgr B2 {and the surrounding environment} to estimate the intensity $I_{6.4} \approx (3-5) \frac{Z}{Z_{\odot}} \times 10^{-6}$\,ph\,cm$^{-2}$\,s$^{-1}$ of {the LECR-induced} Fe K$\alpha$ emission from the Sgr B2 complex. The range of $3-5$ depends on the details of the ambient LECR spectrum and the gas distribution of the cloud.

The {measured} Fe K$\alpha$ flux of ${(10.8\pm1.2)}\times10^{-6}$\,ph\,cm$^{-2}$\,s$^{-1}$ from the envelope (0.5$^{\prime} - 2.2^{\prime}$) region in 2018 (Table~\ref{tab:lecr_regions}) is comparable to the calculation {by \citet{Dogiel15}} if $Z/Z_{\odot} = 2$.
Unless $Z/Z_{\odot} > 2$, ambient LECR protons in this model cannot explain all of the Fe K$\alpha$ emission. However, propagation of ambient LECR protons could contribute  $>$$50\%$ of the 2018 Fe K$\alpha$ emission. 

\citet{Kuznetsova21} disfavor LECR proton propagation as the sole source of the $30-80$\,keV emission from 2011 to 2019 on the basis of the high overall flux compared to the observed hydrogen ionization rate. As detailed in Section~\ref{sec:discussion_XRN}, the $30-80$\,keV band could have {more} substantial contributions from multiple scattering {compared to the Fe K$\alpha$ line}, and a portion of the flux in the 6$^{\prime}$ region of {\em INTEGRAL} is due to X-ray reflection from substructures. {With improved characterization of the flux contribution expected from multiple scattering, enabled by future high-resolution observations of the 6.4\,keV line and the hard X-ray flux as discussed in Section~\ref{sec:discussion_XRN}, the LECR contribution could be more tightly constrained even as it coexists with the emission from multiple scattering.}

Previous work has demonstrated that an ambient population of LECR electrons cannot explain the full hydrogen ionization rate in Sgr B2.
The X-ray continuum from propagation of ambient LECR electrons is expected to have $\Gamma\sim1$, harder than the $\Gamma\sim2$ expected from LECR protons or X-ray reflection and observed from Sgr B2. Given the population of ambient LECR electrons that could explain the observed hydrogen ionization, the hard continuum should have been observable as early as 2009 \citep{Dogiel15,Zhang15,Kuznetsova21}.} 
{While ambient LECR electrons are excluded as the sole origin of hydrogen ionization in Sgr B2, they may still contribute to the nonthermal X-ray emission observed in 2018 or in the future. The sensitivity of the 2018 observations to LECR electrons is restricted by the limitations of the {\em NuSTAR} observation. With spatial resolution for X-rays up to 79\,keV, a future {\em NuSTAR} observation less severely contaminated by stray light could resolve this ambiguity. 

We additionally considered the annihilation or decay of dark matter as a source of the LECRs responsible for the hydrogen ionization and the nonthermal X-ray emission, concluding that any contribution from dark matter would be small compared to the LECR electron population needed to explain the X-ray emission (details in Appendix~\ref{sec:internal_cr}).


\section{Conclusion}\label{sec:conc}

The X-ray features of Sgr B2 provide a window into past energetic activity of Sgr A*, via X-ray reprocessing in the cloud, and to {the GC LECR} populations, via ionization and Bremsstrahlung processes. 

In this paper, we have presented the 103\,ks and 149.2\,ks observations of Sgr B2 taken jointly in 2018 by the {\em XMM-Newton} and {\em NuSTAR} X-ray telescopes, respectively. 
These data show that the {the 2018 Fe K$\alpha$ emission from the central region is $0.8\pm0.2$ of the 2012 level reported by \citet{Zhang15}, consistent with significant contribution from a stationary component. 
Meanwhile, they also reveal new brightening substructures, which, assuming they correspond to clumps within the Sgr B2 complex, are illuminated by one or more secondary external flares.} 
{Based on both the Fe K$\alpha$ light curve and the spectral analysis,} the 2018 {emissions from the central region} are consistent with arising primarily from {X-ray reprocessing}, with possible contributions from the tail of the original flare, multiple scattering albedo, and secondary flares, or {with arising} primarily from LECR interactions. Thus, the flux levels presented {in Table~\ref{tab:lecr_regions}} represent best upper limits on {fluorescence from} LECR interactions within different cloud regions. The Fe K$\alpha$ emission observed from the Sgr B2 envelope is comparable with expectation from the low-energy cosmic proton population that would simultaneously explain {the observed} hydrogen ionization rates in the model of \citet{Dogiel15}.

Future observations of Sgr B2 by {\em XMM-Newton} will {further} constrain the Fe K$\alpha$ light curves from the envelope, core, and the {diffuse substructures}, clarifying the origins of the 2018 emission {and the duration of the external flare(s) illuminating the bright substructures}. 
Meanwhile, further observations {resolving the dense cores can clarify the contribution of multiple scattering and facilitate correspondingly more precise limits on the contributions of LECRs. Two regimes are of particular interest. For hard X-rays above 10\,keV, scattering is the dominant photon process, and a {\em NuSTAR} observation less contaminated by stray light could detect deviations in the light curve of the densest cores relative to the envelope and diffuse cloud regions. For the Fe K$\alpha$ line,  future high-resolution spectrometers could directly resolve the line features, including the Compton shoulder, expected in multiple scattering. }
If further decrease of the Fe K$\alpha$ emission from the envelope is observed, {or if a significant portion of the 2018 Fe K$\alpha$ emission level is definitively attributed to multiple scattering}, the LECR proton model of hydrogen ionization \citep{Dogiel15} {would} begin to be constrained.


\acknowledgments

This research made use of data obtained with {\em XMM-Newton}, an ESA science mission with instruments and contribution directly funded by ESA Member States and NASA. This work is supported by XMM-Newton AO Cycle-16 observation grant 80NSSC18K0623. This research also made use of data from {\em NuSTAR}, a project led by the California Institute of Technology, managed by the Jet Propulsion Laboratory, and funded by NASA. We thank the {\em NuSTAR} Operations, Software and Calibration teams for support with the execution and analysis of these observations. This research has additionally made use of data obtained through the High Energy Astrophysics Science Archive Research Center Online Service, provided by the NASA/Goddard Space Flight Center.

We gratefully acknowledge Chuck Hailey, Roman Krivonos, Ekaterina Kuznetsova, Kaya Mori, Gabriele Ponti, and Farhad Yusef-Zadeh for enlightening discussions related to the Galactic Center cosmic rays and the Sgr B2 region. Special thanks go to Brandon Roach for his data visualization expertise. {We also thank both anonymous reviewers for their feedback, which strengthened and clarified the manuscript.} 

F.R.\ is supported through the National Science Foundation Graduate Research Fellowship Program under Grant No.\ 1122374. M.C.\ acknowledges financial support from the French National Research Agency in the framework of the “Investissements d’avenir” program (ANR-15-IDEX-02) and from CNES.

\vspace{5mm}

\facilities{XMM-Newton {- The X-ray Multi-Mirror Mission}, NuSTAR {- The Nuclear Spectroscopic Telescope Array Mission}}

\software{{XSPEC \citep{Arnaud96},		ESAS \citep{Snowden08},} 		heasoft \citep{Heasoft},		ciaotools \citep{Ciao},		MyTorus \citep{MyTorus}		SAOImage ds9 \citep{ds9},		Matplotlib \citep{matplotlib},		IPython \citep{ipython},		and		Astropy \citep{astropy1,astropy2}
}

\appendix
\section{Spectral Fitting of the Sgr B2 Core Using X-ray Reflection Models}\label{app:xrnspec}

\begin{deluxetable*}{LCCCCChh}
\vspace{5mm}
\tablecaption{Best-fit spectral parameters are shown for a joint fit of the 2018 {\em XMM-Newton} and {\em NuSTAR} observations, using the central 90$^{\prime\prime}$ of Sgr B2 for {\em XMM-Newton} and central 50$^{\prime\prime}$ of Sgr B2 for {\em NuSTAR}. We report flux parameters for the 90$^{\prime\prime}$ region. \label{tab:paramsApp}}
\tablewidth{0pt}
\tablehead{
\colhead{Parameter} &\colhead{Unit}&\colhead{Phenomenological\tablenotemark{a}} &\colhead{MyTorus\tablenotemark{b}} &\colhead{Walls\tablenotemark{c}} &\colhead{CREFL16\tablenotemark{d}} &\nocolhead{LECRe\tablenotemark{e}} &\nocolhead{LECRp\tablenotemark{e}} 
}
\startdata
N_H(f)& 10^{23}\,\textrm{cm}^{-2}&{ 0.9^{+0.2}_{-0.1}}&1.4\pm0.1&0.9^{+0.1}_{-0.2}&1.3\pm0.1&0.9\pm0.1&0.9\pm0.1\\
N_H(i)& 10^{23}\,\textrm{cm}^{-2}&{4.6^{+0.7}_{-0.6}}&9.0^{+6.0}_{-2.5}&7.9^{+3.7}_{-2.1}&12.3^{+7.9}_{-4.5}&5.2^{+1.2}_{-1.1}&5.0^{+0.4}_{-1.0}\\
Z/Z_\odot\textrm{ (apec)}& &2\textrm{*}&2\textrm{*}&2\textrm{*}&2\textrm{*}&2\textrm{*}&2\textrm{*}\\
Z/Z_\odot \textrm{ (cloud)}& &...&...&...&1.0\textrm{*}&1.9^{+0.8}_{-0.4}&0.5^{+0.3}_{-...}\\
kT& \textrm{keV}&{4.3^{+1.0}_{-0.7}}&6.5^{+0.8}_{-0.7}&4.3^{+1.1}_{-0.8}&5.4^{+0.6}_{-0.9}&4.3^{+1.1}_{-0.7}&4.3^{+1.1}_{-0.7}\\
F_{apec}~(2-10\textrm{\,keV})& 10^{-13} \textrm{\,erg\,cm}^{-2}\textrm{\,s}^{-1}&{5.6\pm0.3}&9.5\pm0.4&5.7\pm0.3&7.6^{+1.2}_{-0.9}&5.6\pm0.3&5.6\pm0.3\\
F_{6.4\,keV}& 10^{-6}\textrm{\,ph\,cm}^{-2}\textrm{\,s}^{-1}&{6.7\pm0.8}&...&...&...&...&...\\
\Gamma_{pl}& &2.0\textrm{*}&...&...&...&...&...\\
\theta_{XRN}& &...&...&28^{+10}_{-16}&64^{\circ}\textrm{*}&...&...\\
\Gamma_{XRN}& &...&2.6\pm...&1.8^{+0.6}_{-...}&2.5^{+...}_{-0.6}&...&...\\
\textrm{{multiplicative} factor}& &{0.11^{+0.04}_{-0.03}}&0.19\pm0.06&0.11^{+0.02}_{-0.03}&0.15^{+0.02}_{-0.04}&0.12^{+0.06}_{-0.04}&0.12^{+0.06}_{-0.04}\\
\chi^2_{\nu}~\textrm{(d.o.f)}& &{1.07~(257)}&1.29~(257)&1.09~(256)&1.16~(257)&1.08~(255)&1.08~(255)\\
\enddata
\tablecomments{The goodness of fit is estimated by $\chi^2_{\nu}$ and the number of degrees of freedom is given in parentheses. The errors represent 90\% confidence. The fluxes and normalizations are for the 90$^{\prime\prime}$ region. The {multiplicative} factor relates the flux from the 50$^{\prime\prime}$ {\em NuSTAR} source region to the 90$^{\prime\prime}$ {\em XMM-Newton} region. 
}
\tablenotetext{a}{Fitting with the phenomenological model ({\tt wabs*(apec+wabs*po+ga+ga)} in {\tt XSPEC}) is detailed in Section~\ref{sec:pospec} and Table~\ref{tab:params}. The model parameters are repeated here for comparison with the XRN spectral models.}
\tablenotetext{b}{The MyTorus  \citep{Murphy09} model is given by {\tt wabs*(apec + MYTS + MYTL)} in {\tt XSPEC}, where {\tt MYTS} and {\tt MYTL} are two models representing the scattered continuum and fluorescent line emissions, respectively. {\tt MyTorus} is parameterized by the photon index $\Gamma$ of the external X-ray population, the internal column density $N_{H}(i)$, and an overall normalization $N$, with all parameters coupled between {\tt MYTS} and {\tt MYTL} for consistency.}
\tablenotetext{c}{The Walls model is given by {\tt wabs*(apec+Walls)} in {\tt XSPEC} \citep{Walls16}, where {\tt Walls} is parametrized by the photon index $\Gamma$ of the external X-ray population, the internal column density $N_H(i)$, and the inclination angle $\theta$ of the cloud relative to the X-ray source, as well as an overall normalization factor $N$.}
\tablenotetext{d}{The CREFL16 model is given by {\tt wabs*(apec+CREFL)} in {\tt XSPEC} \citep{Churazov17}.  It is parametrized by metallicity $Z/Z_{\odot}$, $\theta$, $\Gamma_{XRN}$, and $\tau_T= 2\sigma_T N_H(i)$, where the Thomson cross section {$\sigma_T = 6.65 \times10^{-25}$\,cm$^2$}, as discussed in the text. Here we report $N_H(i)$ directly for comparison with the other models.}
\tablenotetext{*}{Starred parameters were not allowed to vary in the fit.}
\end{deluxetable*}

{This appendix details the spectral fitting of the 2018 observations with the three self-consistent XRN spectral models in Section~\ref{sec:xrnspec}. These fits show that the 2018 spectra are not inconsistent with arising primarily from X-ray reflection.
Table~\ref{tab:paramsApp} summarizes the fitting results.} 

Previously, \citet{Zhang15} modeled X-ray reflection from Sgr B2 using the {\tt MyTorus} X-ray Reprocessing Model \citep{Murphy09, MyTorus} developed for X-ray reflection from toroid structures surrounding Compton-thick AGN. 
{\tt MyTorus} is parametrized by the cloud hydrogen column density $N_H(i)$, the spectral index $\Gamma$ of the external X-ray source, and an overall normalization. 
The model is additionally parametrized by an inclination angle of the torus relative to the source, which we fix at 0$^{\circ}$ to most closely describe the situation of a spherical cloud. 
{\tt MYTorus} has additional limitations beyond geometry, detailed by \citet{Zhang15}, when applied to model MCs as XRN, and in particular it assumes reflection in a medium with metallicity $Z = Z_{\odot}$ and uniform density. 
{\tt MYTorus} is implemented as a series of models in {\tt XSPEC}. To best approximate the MC scenario, we used two {\tt MyTorus} components, describing the scattered continuum ({\tt MYTS}) and the iron fluorescence lines ({\tt MYTL}), with all parameters coupled between these components so that they were consistent with arising from the same incident X-rays. 
For consistency with \citet{Zhang15}, we used the models with reflected X-rays up to 500\,keV and an energy offset of $+40$\,eV for the fluorescent lines. Thus the {\tt XSPEC} model is {\tt wabs*(apec+MYTS+MYTL)}, where the
MYTorus components are {\tt mytorus\_scatteredH500\_v00.fits} and {\tt mytl\_V000010pEp040H500\_v00.fits}, and the thermal plasma {\tt apec} parameters are as in Section~\ref{sec:pospec}.
The fit quality is marginally acceptable with $\chi^2_{\nu} = 1.29$ for 257 d.o.f. Additionally, both the intrinsic and foreground column densities are fitted much higher than the physical expectation, and the {\tt apec} component absorbed a flux somewhat larger than expected based on the phenomenological model fit, possibly because it was compensating for low compatibility of the {\tt myTorus} model with the data. Thus, the 2018 data are not well suited for the {\tt MyTorus} model with physical parameters. However, due to the limitations of the {\tt myTorus} model in describing Sgr B2, we do not exclude the XRN scenario based on this result.

The Monte Carlo model developed by \citet{Walls16}, hereafter the {\tt Walls} model, was developed specifically to model MCs as XRN with the particular application of constraining the location of Sgr B2 relative to Sgr A*. Like {\tt MyTorus}, {\tt Walls} depends on intrinsic absorption $N_H(i)$ of the cloud and spectral index $\Gamma$ of the external X-ray source. It is additionally parametrized by the angle $\theta$ of the spherical cloud relative to the external source, such that $\theta = 0$, which is unphysical given the non-zero projected distance from Sgr B2 to Sgr A*, corresponds to the cloud situated in line between the source and the observer, and the true distance from the X-ray source to the cloud is $\sim$100$\textrm{\,pc}/\sin\theta$. The position $\theta$ of the cloud relative to the source strongly impacts the XRN spectrum, as absorption patterns and effective cloud thickness differ dramatically between X-rays reflected to observers at different $\theta$. 
There are several {\tt Walls} models with varying cloud metallicities and density profiles. We chose the model with $Z = Z_{\odot}$ for direct comparison with spectral fitting by \citet{Walls16} but note that with these data the metallicity does not significantly change the spectrum below 10\,keV, apart from the overall normalization of incident X-rays. We used a uniform, rather than Gaussian, cloud density profile, which gave a better fit and was found to be more suitable by \citet{Walls16}. Thus the model is {\tt wabs*(apec+Walls)} where the {\tt Walls} component is {\tt Walls\_et\_al\_2016\_Uniform\_1.0.fits}. 

The fit was satisfactory, with $\chi^2_{\nu}$ = 1.09 for 256 d.o.f. We found the best fit $\theta = 28^{+10}_{-16}$, significantly lower than the best fit of $\theta = 64^{+8}_{-7}$ found for the bright state \citep{Walls16}. However, since these data do not represent X-ray reflection in the bright state, we caution against overinterpretation of the result. The foreground column density was consistent with previous results at $N_H(f) = 0.9^{+0.1}_{-0.2} \times 10^{23}\textrm{\,cm}^{-2}$, and the fitted intrinsic column density was on the high side of previous results, at $N_H(f) = 7.9^{+3.7}_{-2.1} \times 10^{23}\textrm{\,cm}^{-2}$. While the source XRN spectral index was not constrained, the flux and temperature of the {\tt apec} component were consistent with expectation, and the data are consistent with XRN origin according to the {\tt Walls} model.

Finally, the uniform Cloud REFLection of 2016, {\tt CREFL16}, model \citep{Churazov17} treats the case of XRN in a uniform spherical cloud using the same geometry as \citet{Walls16} but with additional fluorescence lines (beyond Fe) and treating metallicity $Z/Z_{\odot}$ as a free parameter. In addition to $Z/Z_{\odot}$, the model parameters include the inclination $\cos\theta$ to the illuminating source, the spectral index $\Gamma$ of the illuminating source spectrum, and the optical depth of Thomson scattering, $\tau_{T} = \sigma_{T}n_HR$ where {$\sigma_{T} \sim 6.65\times10^{-25}$\,cm$^2$} is the Thomson scattering cross section, $n_H$ is the hydrogen density, and R is the cloud radius, such that $N_H (i) = 2n_HR = 2\tau_T/\sigma_T$. In Table~\ref{tab:params} we report $\theta$ and $N_H(i)$ for direct comparison with other models. Because the {\tt CREFL16} fit favored the unphysical result $\theta = 0$, we fixed $\theta = 64^{\circ}$ as in \citet{Walls16}. Because the best fit gave $Z/Z_{\odot} < 1$, which is unphysical, we froze the metallicity at solar abundance, the lowest abundance consistent with observation. The fit was acceptable with $\chi^2_{\nu} = 1.16$ for 257 dof., but the {\tt apec} component absorbed an order of magnitude more flux than expected, suggesting as with {\tt Walls} that the {\tt CREFL16} model is not best suited for these data. Meanwhile, the {\tt apec} and absorption parameters were consistent with those in the {\tt Walls} fit. 

For all three self-consistent XRN models, $\chi_{\nu}^2$ was acceptable, but most model parameters were poorly constrained. In the {\tt CREFL16} model, the best fit angle to Sgr B2 was $\theta = 0$, but the fit was still acceptable with the constraint $\theta = 64^{\circ}$. While $\theta = 0$ does not reflect the position of Sgr B2, the preference for $\theta = 0$ may reflect a situation in which X-rays observed from Sgr B2 in 2018 are produced after the external X-ray front has partially or fully passed through the cloud. In this case, the external X-rays still being reflected have already traversed a cloud depth larger than the typical reflection depth in the bright state, and the effects of multiple scattering may also become important. The fits in this section illustrate that the 2018 flux from the core of Sgr B2 can be described by the XRN spectral models. However, even if X-ray reflection dominates the 2018 flux, the XRN models do not account for time-dependent effects as the external front passes through the cloud, so we cannot use these results to constrain the external X-ray spectrum.

\section{Spectral Fitting of Sgr B2 Regions}\label{app:lecr_region_fits}
Table~\ref{tab:regions} lists the sky regions used in the {analysis} of {1) }time variability of the Sgr B2 structures in Section~\ref{sec:timing} {(fitting details in Section~\ref{sec:timingdet})} as well as {2)}  LECRs in different portions of the cloud in Section~\ref{sec:lecr} {(fitting details in Section~\ref{sec:lecrdet})}. Circular and annular regions centered on the Sgr B2 core are not listed in the table, but are centered at RA=17$^{\textrm{h}}$47$^{\textrm{m}}$19.992$^{\textrm{s}}$, Dec=$-$28$^{\circ}$23$^{\prime}$07.08$^{\prime\prime}$.

\subsection{Analysis of Sgr B2 Time Variability}\label{sec:timingdet}
{The light curves in Section~\ref{sec:timing} (Figures~\ref{fig:cloudDecrease} and \ref{fig:substructures}) were produced using spectral fits to data extracted from four different {\em XMM-Newton} observations, extracted using a local background region selected for each observation. The source regions, and the background region for each observation, are given in Table~\ref{tab:regions}.} 
To produce each light curve, we first selected the observation when that substructure was brightest. {If the 2001 observation was brightest, as for the regions in Figure~\ref{fig:cloudDecrease}, we selected the 2004 observation due to the low statistics of the 2001 observation. We} performed a spectral fit to the phenomenological model {as} in Section~\ref{sec:pospec}, {except that we froze} $N_H(f) = {0.9\times} 10^{23}\,\textrm{cm}^{-2}$ for consistency between regions {and there was no multiplicative factor in the absence of {\em NuSTAR} data}. 
Because the structures are too dim in some epochs to constrain a good fit, we then fixed all model parameters other than the normalizations before fitting the spectra from the other observations. 
Though constraints based on the large size of Sgr B2, hard point sources, and different focused fields of view necessitated use of different sky regions for background subtraction in each observation {(see Table~\ref{tab:regions}), cross checks using different background regions in the same observation yielded consistent Fe K$\alpha$ flux measurements. }

\subsection{LECR Analysis}\label{sec:lecrdet}
Figure~\ref{fig:lecr_fits} shows the spectral fits with the phenomenological model {used to obtain the Fe K$\alpha$ fluxes reported} in Table~\ref{tab:lecr_regions}. 
Because of the restricted {\em NuSTAR} field of view, these data were based on fits to the 2018 {\em XMM-Newton} observations only, 
{with spectra extracted using the source and background regions in Table~\ref{tab:regions}.}
Spectral fitting with the phenomenological model {is} as in Section~\ref{sec:pospec}, except that we froze $N_H({f}) = {0.9\times}10^{23}$\,cm$^{-2}$ for a fair comparison between regions {and there was no multiplicative factor in the absence of {\em NuSTAR} data}. 
The resulting fits were satisfactory as evaluated by $\chi^2_{\nu}$. 
{The Fe K$\alpha$ flux was evaluated using the {\tt cflux} command in {\tt XSPEC}}. 
{While the choice of $N_H(f)$ introduces a systematic uncertainty into the total Fe K$\alpha$ flux calculation, it is expected to be small compared to the statistical uncertainty, as a $10\%$ change in $N_H(f)$ produces a $\sim$$1.5\%$ change in the fitted line flux at 6.4\,keV.}  
The surface brightness {in Table~\ref{tab:lecr_regions}} was calculated using the region area obtained by image analysis (i.e.\ accounting for missing pixels).

\begin{deluxetable*}{lccCCCc}
\vspace{5mm}
\tablecaption{Sky regions used for spectral extraction in Section~\ref{sec:timing} (upper), in Section~\ref{sec:lecr} (middle), and for local background subtraction throughout this work (lower). }
\tablewidth{0pt}
\tablehead{
\colhead{Region Name} &\colhead{R.A.}&\colhead{Dec.} &\colhead{Radius}&\colhead{Minor axis} &\colhead{Major axis} &\colhead{Angle} 
}
\startdata
G0.74-0.10 &   17$^{\textrm{h}}$47$^{\textrm{m}}$44.878$^{\textrm{s}}$& 		$-$28$^{\circ}$21$^{\prime}$22.60$^{\prime\prime}$ 		&...&60$^{\prime\prime}$   & 	150$^{\prime\prime}$ & 91.212$^{\circ}$ \\
G0.75-0.01 &   17$^{\textrm{h}}$47$^{\textrm{m}}$22.412$^{\textrm{s}}$& 		$-$28$^{\circ}$18$^{\prime}$38.91$^{\prime\prime}$& 	40$^{\prime\prime}$   & ...& ... &...\\
G0.66-0.13 &   17$^{\textrm{h}}$47$^{\textrm{m}}$41.950$^{\textrm{s}}$&	 	$-$28$^{\circ}$26$^{\prime}$23.15$^{\prime\prime}$& 	...&72$^{\prime\prime}$   &144$^{\prime\prime}$ & 121.212$^{\circ}$ \\
G0.66-0.13 A &17$^{\textrm{h}}$47$^{\textrm{m}}$39.737$^{\textrm{s}}$&		$-$28$^{\circ}$24$^{\prime}$58.48$^{\prime\prime}$& 	40$^{\prime\prime}$   & ...&...&... \\
G0.66-0.13 B &17$^{\textrm{h}}$47$^{\textrm{m}}$43.860$^{\textrm{s}}$& 		$-$28$^{\circ}$27$^{\prime}$08.63$^{\prime\prime}$& 	40$^{\prime\prime}$   & ...&...&... \\
G0.66-0.13 C &17$^{\textrm{h}}$47$^{\textrm{m}}$38.365$^{\textrm{s}}$&		$-$28$^{\circ}$25$^{\prime}$45.64$^{\prime\prime}$& 	40$^{\prime\prime}$   & ...&...&... \\
G0.56-0.11 &   17$^{\textrm{h}}$47$^{\textrm{m}}$24.879$^{\textrm{s}}$&		$-$28$^{\circ}$30$^{\prime}$50.91$^{\prime\prime}$& 	 90$^{\prime\prime} $  & ... & ...&... \\
G0.56-0.11 A &17$^{\textrm{h}}$47$^{\textrm{m}}$27.374$^{\textrm{s}}$&		$-$28$^{\circ}$29$^{\prime}$30.58$^{\prime\prime}$ & 	40$^{\prime\prime} $  & ...&... &...\\
G0.56-0.11 B&17$^{\textrm{h}}$47$^{\textrm{m}}$24.053$^{\textrm{s}}$&		$-$28$^{\circ}$30$^{\prime}$52.82$^{\prime\prime}$& 	40$^{\prime\prime}$   & ...&... &...\\
G0.56-0.11 C& 17$^{\textrm{h}}$47$^{\textrm{m}}$15.249$^{\textrm{s}}$&		$-$28$^{\circ}$31$^{\prime}$2712$^{\prime\prime}$& 	40$^{\prime\prime}$   & ...&... &...\\
G0.56-0.11 D& 17$^{\textrm{h}}$47$^{\textrm{m}}$18.848$^{\textrm{s}}$&		$-$28$^{\circ}$32$^{\prime}$57.32$^{\prime\prime}$& 	40$^{\prime\prime} $  & ...&... &...\\
G0.61+0.00 &  17$^{\textrm{h}}$47$^{\textrm{m}}$03.925$^{\textrm{s}}$&		$-$28$^{\circ}$24$^{\prime}$54.11$^{\prime\prime}$& 	...&72$^{\prime\prime}$    &144$^{\prime\prime}$  &  16.203$^{\circ}$ \\
G0.61+0.00 A &17$^{\textrm{h}}$47$^{\textrm{m}}$04.276$^{\textrm{s}}$&		 $-$28$^{\circ}$24$^{\prime}$40.44$^{\prime\prime}$& 	40$^{\prime\prime}$   & ...&...&... \\
G0.61+0.00 B& 17$^{\textrm{h}}$47$^{\textrm{m}}$06.275$^{\textrm{s}}$&		$-$28$^{\circ}$26$^{\prime}$44.07$^{\prime\prime}$ &40$^{\prime\prime}$   & ...& ...&...\\
6$^{\prime}$ Region    &17$^{\textrm{h}}$47$^{\textrm{m}}$29.280$^{\textrm{s}}$&		$-$28$^{\circ}$21$^{\prime}$57.60$^{\prime\prime}$ &360$^{\prime\prime} $  & ...& ...&...\\
\hline
Diffuse Ellipse & 17$^{\textrm{h}}$47$^{\textrm{m}}$37.123$^{\textrm{s}}$&	$-$28$^{\circ}$17$^{\prime}$26.16$^{\prime\prime}$& 	...& 104$^{\prime\prime} $ & 236$^{\prime\prime}$ & 121.354$^{\circ}$ \\
Envelope Ellipse & 17$^{\textrm{h}}$47$^{\textrm{m}}$18.070$^{\textrm{s}}$&	$-$28$^{\circ}$21$^{\prime}$24.22$^{\prime\prime}$& ...&36$^{\prime\prime}$  & 102$^{\prime\prime} $& 255$^{\circ}$ \\
\hline
{{\em NuSTAR} Background} & 17$^{\textrm{h}}$47$^{\textrm{m}}$17.695$^{\textrm{s}}$&	$-$28$^{\circ}$27$^{\prime}$09.13$^{\prime\prime}$& 	...& 65$^{\prime\prime} $ & 125$^{\prime\prime}$ & 93$^{\circ}$ \\
{{\em XMM-Newton} Background A}\textrm{*} & 17$^{\textrm{h}}$46$^{\textrm{m}}$45.638$^{\textrm{s}}$&	$-$28$^{\circ}$13$^{\prime}$29.32$^{\prime\prime}$& ...&115$^{\prime\prime}$  & 150$^{\prime\prime} $& 85$^{\circ}$ \\
{{\em XMM-Newton} Background B\textrm{*}} & 17$^{\textrm{h}}$48$^{\textrm{m}}$23.336$^{\textrm{s}}$&	$-$28$^{\circ}$32$^{\prime}$27.73$^{\prime\prime}$& ...&115$^{\prime\prime}$  & 150$^{\prime\prime} $& 30$^{\circ}$ \\
{{\em XMM-Newton} Background C\textrm{*}} & 17$^{\textrm{h}}$46$^{\textrm{m}}$19.036$^{\textrm{s}}$&	$-$28$^{\circ}$27$^{\prime}$40.30$^{\prime\prime}$& ...&115$^{\prime\prime}$  & 150$^{\prime\prime} $& 20$^{\circ}$ \\
\enddata
\tablecomments{All coordinates are in terms of right ascension (R.A.) and declination (Dec.) using the J2000 system. {Circular and annular regions centered on the Sgr B2 core are not listed in the table but} assume the Sgr B2 complex is centered on RA=17$^{\textrm{h}}$47$^{\textrm{m}}$19.992$^{\textrm{s}}$, Dec=$-$28$^{\circ}$23$^{\prime}$07.08$^{\prime\prime}$.\label{tab:regions}
}
 \tablenotetext{*}{No single suitable region located outside of the spatial extent of Sgr B2 and unaffected by hard point sources was also compatible with the field of view of all four {\em XMM-Newton} observations. Thus, multiple background regions were used. {\em XMM-Newton Background  A} was used for all analyses with the 2018 data (0802410101). The alternate {\em XMM-Newton Background B} was used for the 2012 and 2004 observations (0694640601 and 0203930101) while {\em XMM-Newton Background C} was used for the 2001 observation (0112971501).}
\end{deluxetable*}

\begin{figure*}
\fig{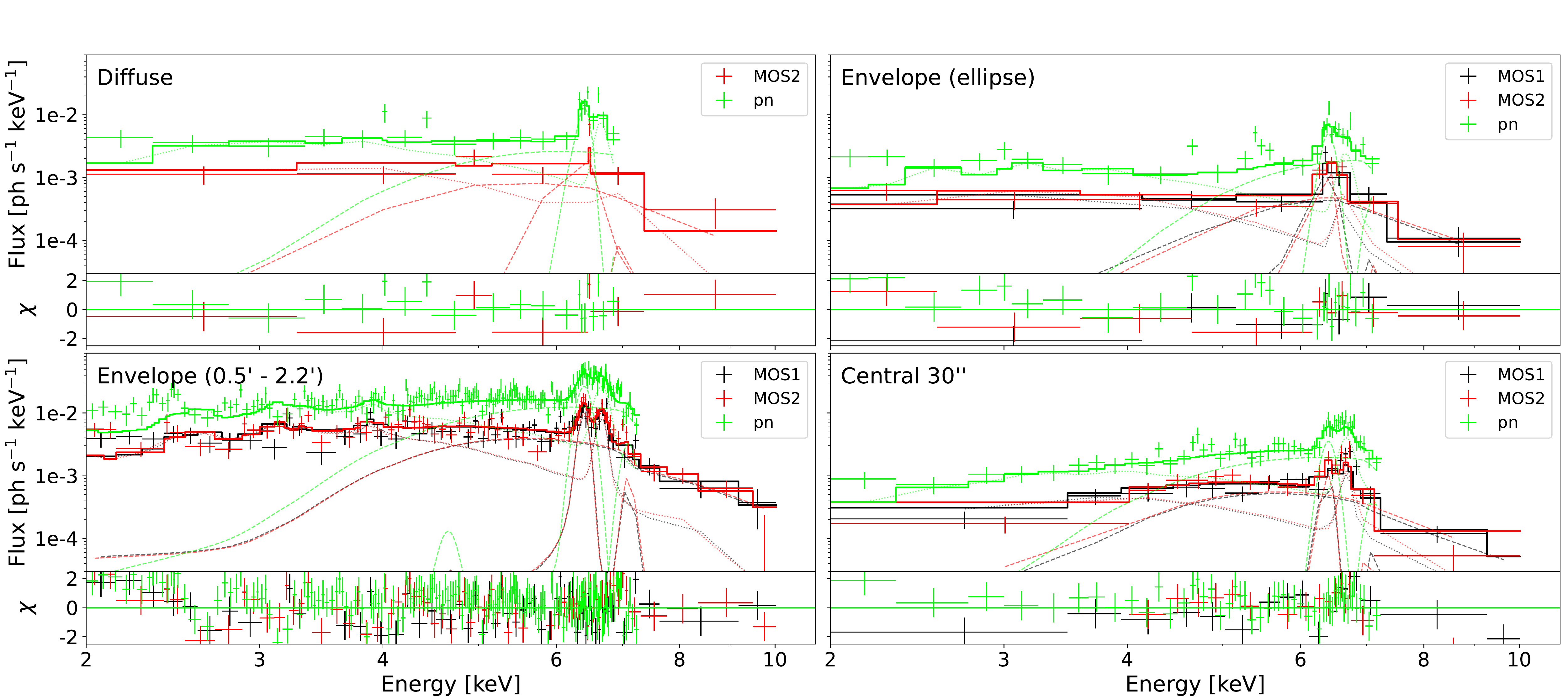}{\textwidth}{}
\caption{
The 2018 spectral fits that yield the data presented in Table~\ref{tab:lecr_regions}, are shown. {Background-subtracted} spectra are shown for the Diffuse (upper left), Envelope (ellipse, upper right),  Envelope ($0.5^{\prime}-22^{\prime}$, lower left), and Core ($0.5^{\prime}$, lower right) regions (given in Figure~\ref{fig:nH}), fitted with the phenomenological model. The models are fitted simultaneously to data from the {\em XMM-Newton MOS1} (black, $2-10$\,keV) {\em MOS2} (red, $2-10$\,keV) and {\em pn} (green, $2-7.8$\,keV). The Diffuse region is not compatible with the {\em MOS1} field of view. 
The data are binned with $3\sigma$ confidence. The best fit is shown in the horizontal lines. 
{The contributions of the {\tt apec} (dotted) and the nonthermal spectral components (dashed; {\tt ga} and {\tt po}) are also shown. Deviations between the data and the model are most pronounced at low energies.}
\label{fig:lecr_fits}}
\end{figure*}

\section{Limits on Dark Matter Annihilation}\label{sec:internal_cr}

The annihilation of dark matter has been of interest as a possible source of GC CRs (\citet{Murgia20} and references therein) including GC LECRs (e.g.\ \citet{Linden11}).
Here we consider the case of LECRs produced within Sgr B2 in the annihilation of dark matter as the nonthermal source responsible for the observed Fe K$\alpha$ emission. 

Any model of DM annihilating to low-energy electrons or protons would produce a LECR population within GC MCs and thus induce Fe K$\alpha$ fluorescence. In  contrast to the ambient LECR scenario of Section~\ref{sec:ambient_cr}, LECR illumination {would be expected} throughout the Sgr B2 core and envelope. 
{Considering the production of LECR from dark matter,} the generic dark matter annihilation rate per unit volume is:
\begin{equation}\label{eq:annihilation}
\Phi_{DM} = \frac{1}{2}\langle\sigma v\rangle\left(\frac{\rho}{M_{DM}}\right)^2,
\end{equation}
where the thermally averaged dark matter annihilation cross section $\langle\sigma v\rangle$ and dark matter mass $M_{DM}$ parametrize the model, and the local dark matter energy density $\rho$ is estimated as $\rho \sim100$\,GeV/cm$^3$ in Sgr B2, given $\sim$$100$\,pc distance from the GC \citep{Linden11}.

To estimate an overly optimistic scenario for the dark matter origin of LECR electrons with GC MCs, we assumed that all annihilation energy is transferred to electron kinetic energy. We also assume that all electrons produced in the core and envelope are absorbed within those regions, which is reasonable considering the diffusive dynamics of LECRs within the envelope of Sgr B2 in Section~\ref{sec:ambient_cr} and is explicitly treated by \citet{Gabici13}. We considered the 90$^{\prime\prime}$ sky region surrounding the core of Sgr B2, which represents a spherical volume of $\sim$$165$\,pc$^3$ given a distance to Sgr B2 of $\sim$$7.9$\,kpc.
Comparing our fitted maximum power of LECR electrons in the central 90$^{\prime\prime}$ of Sgr B2, $\textrm{d}W/\textrm{d}t < 2.2 \times 10^{40}$, to the annihilation rate in Eq.~\eqref{eq:annihilation}, we calculated { that a dark matter model with} $\langle\sigma v\rangle$ larger than $3\times10^{-22}$\,cm$^3$\,s$^{-1}$ ($M_{DM}\sim 500$\,keV) to 6$\times10^{-19}$\,cm$^3$\,s$^{-1}$ ($M_{DM}\sim1$\,GeV) {is necessary to produce a LECR electron population sufficient to explain the bulk of the Fe K$\alpha$ emission observed from Sgr B2}.

This {minimum possible $\langle\sigma v\rangle$} is several orders of magnitude larger than both the existing best limits on $\langle\sigma v\rangle$ for weakly interacting massive particle (WIMP) dark matter in this mass range annihilating to all visible products and the thermal relic annihilation cross section predicted by theory \citep{NotEvenSlightlyDead}. 
Because this calculation assumes that all dark matter annihilation energy is transferred to the cloud via ionization and excitation by CRs, it is perfectly general to the case of hidden sector dark matter models as well. As with the WIMP case, existing limits on $\langle\sigma v\rangle$ for hidden sector dark matter \citep{Elor16} are several orders of magnitude smaller than compared to a dark matter model necessary to explain the fluorescence in Sgr B2.
Thus, annihilation of dark matter cannot produce a LECR electron population capable of producing the observed Fe K$\alpha$ fluorescence in Sgr B2.

\clearpage

\bibliography{../../../main}{}
\bibliographystyle{aasjournal}

\end{document}